\documentclass[12pt,single]{article} 
\usepackage{anysize}

\usepackage[dvips]{graphicx} 
\usepackage{amsmath}
\usepackage{amssymb} 
\usepackage{amsfonts}
\usepackage[hang,small,bf]{caption}

\begin{document}

\begin{center}
\section*{Lamb   Shift in Light Muonic Atoms  - Revisited}
\end{center}    

\subsection*{E. Borie}

\subsubsection*{Karlsruhe Institute of Technology,   \\                                         
Institut f\"ur Hochleistungsimpuls and Mikrowellentechnik (IHM), \\ 
Hermann-von-Helmholtzplatz~1,\\ D-76344 Eggenstein-Leopoldshafen, Germany}


\vspace{1.5cm}

\subsubsection*{Abstract}
\vspace{-0.2cm}
In connection with recent and proposed experiments, 
and new theoretical results, my  previous calculations of the Lamb shift
in muonic hydrogen will be reviewed and compared with other work.  In
addition, numerical results for muonic deuterium and helium  will be presented.
Some previously neglected (but very small) effects are included.

\subsubsection* { Introduction }  
\vspace{-0.2cm}

The energy levels of muonic atoms are very sensitive to effects of
quantum electrodynamics (QED), nuclear structure, and recoil, since the
muon is about 206 times heavier than the electron \cite{RMP}. 

 A recent measurement of the Lamb shift in muonic hydrogen \cite{experiment}
 has stimulated great renewed interest in the energy levels of muonic
atoms.   A number of theoretical analyses of the Lamb shift (the 2p-2s
transition) in light muonic atoms have been published  
\cite{eides,Pachuki1,Pachuki2,Borie05,Borie75,Borie78,hyperfine,BorieHe3,digiacomo,carboni}
before the experiment was performed.  The present paper repeats the
independent recalculation of some of the most important effects
\cite{Borie05} for hydrogen and deuterium \cite{Borie-d05}, 
including some previously neglected (but very small) effects and extends
the numerical calculations to the case of muonic helium.
Some recent results by other authors are also discussed.

In the numerical calculations the fundamental constants from the most
recent \\ CODATA compilation  (\cite{codat06})  are used (with the
exception of the fine structure constant, which was revised more
recently), i.e.: \\ 
$\alpha^{-1}$, $\hbar c$, $m_{\mu}$, $m_e$, $m_u$\,=\,137.0359991,
 197.32696\,MeV$\cdot$fm, 
 105.658367\,MeV, 0.5109989\,MeV,  931.494030\,MeV, respectively.     


The changes in these constants compared with  CODATA 2002 (\cite{codat02})
are too small to make any relevant difference in the results, but the
most recent values are used anyway. 

Also, the following properties of the proton and deuteron were used: \\
$m_p$\,=\,938.27203\,MeV/c$^2$,  $\mu_p\,=\,2.79285\,\mu_N$, and
$R_p$\,=\,0.875(7)\,fm (from \cite{codat02}); this agrees with a recent
determination by the group at Jefferson Laboratory \cite{jlab}.  Other
recent values have been reported, for example 0.8418(7)\,fm\,\cite{experiment} 
and 0.877(8)\,fm\,\cite{mainz2010}.  
In some cases calculations
were performed for several values of $R_p$.   In addition (see \cite{codat06}) 
 $m_d$\,=\,1875.613\,MeV/c$^2$, $\mu_d\,=\,0.85744\,\mu_N$,   
$R_d$\,=\,2.139(3)\,fm (from \cite{codat02}), or
$R_d$\,=\,2.130(3)\,fm, (\cite{sick98}). 
For the helium isotopes  
 $m_{\alpha}$\,=\,3727.379\,MeV/c$^2$,   $R_{\alpha}$\,=\,1.676(8)\,fm 
 (\cite{sick82}), or  
 $R_{\alpha}$\,=\,1.681(4)\,fm  (\cite{sick08}); 
 $m_{He3}$\,=\,2808.391\,MeV/c$^2$ and $\mu_{He3}\,=\,-2.1275\,\mu_N$.  
An old measurement of $R_{He3}$\,=\,1.844\,fm \cite{oldHe3} has been
superseded by newer results which range from 1.959(34)\,fm~\cite{sickreview08}
or 1.961\,fm~\cite{newerHe3}  to 1.975(4)\,fm~\cite{newestHe3} (other results 
are cited in these papers).  Here an average value of 1.966(10)\,fm will be 
used for an estimate, since there is still some uncertainty in the calculations 
used in the newer determinations.  In some cases, older values for the proton 
and deuteron radii were used to make comparison with earlier results 
(\cite{Borie05,Borie-d05} easier.   

The deuteron has spin 1 and thus has both magnetic and
quadrupole moments.  The quadrupole moment of the deuteron is taken 
to be Q\,=\,0.2860(15)\,fm$^2$~\cite{friar02,reid-v,bishop}.  
A newer result \cite{newQ} is nearly the same, but more precise.

\subsubsection*{Vacuum Polarization }
\vspace{-0.2cm}
The most important QED effect for muonic atoms is the virtual production
and annihilation of a single $e^+ e^-$ pair.   It has as a
consequence an effective interaction of order $\alpha Z \alpha$ which
is usually called the Uehling potential (\cite{Uehling,serber}.  This
interaction describes the most important modification of Coulomb's law.
Numerically it is so important that it should not be treated using
perturbation theory;  instead the Uehling potential should be added to
the nuclear electrostatic potential before solving the Dirac equation.
However, a perturbative treatment is also useful in the case of very
light atoms, such as hydrogen.  

Unlike some other authors, we prefer to  use relativistic (Dirac) wave
functions to describe  the muonic orbit.  This is more exact, and as
will be seen below, it makes a difference at least for the most
important contributions.  Relativistic recoil corrections to this
treatment have been calculated according to the prescription given in
\cite{RMP}. More recent results using another method \cite{Karshenboim2012} 
will also be given.  The wave functions are given in the book of 
Akhiezer and Berestetskii \cite{akhiezer} and will not be given here.  
In perturbation theory, the energy shift due to an effective potential 
$\Delta V$ is given by 
\begin{equation}
 \Delta E_{n \kappa}~=~ \frac{1}{2 \pi^2} \cdot \int_0^{\infty} q^2 dq
 \Delta V(q) \cdot \int_0^{\infty} dr j_0(qr) [F^2_{n \kappa}+G^2_{n \kappa}]
\label{eq:pert1}
\end{equation}
where $F_{n \kappa}$ and $G_{n \kappa}$ are the small and large components
of the wave function, $n$ is the principle quantum number and $\kappa$
is equal to $-\ell-1$ if $j=\ell+\frac{1}{2}$ and $+\ell$ if
$j=\ell-\frac{1}{2}$.    
$\Delta V(q)$ is the Fourier transform of the physical potential.
\begin{equation}
 \Delta V(q)~=~ 4 \pi \cdot \int_0^{\infty} r^2  \cdot  j_0(qr) 
 \cdot \Delta V(r) \,  dr
\label{eq:dv2}
\end{equation}
\begin{equation}
 \Delta V(r)~=~ \frac{1}{2 \pi^2} \cdot \int_0^{\infty} q^2  \cdot  j_0(qr) 
  \cdot \Delta V(q) \, dq
\label{eq:dv1}
\end{equation}
 As is well-known \cite{RMP}, the Uehling potential in momentum space is
 given by  
\begin{equation*}
 V_{Uehl}(q) \,=\, -\frac{4 \alpha(\alpha Z)}{3} \cdot G_E(q) \cdot  
  F(\phi) \,=\, -4 \pi (\alpha Z) \cdot G_E(q) \cdot U_2(q)
\end{equation*}
\noindent where   $G_E$ is the proton charge form factor, 
$\sinh(\phi) = q/(2 m_e)$ and   
\begin{equation}%
 F(\phi) \,=\, \frac{1}{3} + (\coth^2(\phi)-3) \cdot 
  [1 - \phi \cdot \coth(\phi)]
\label{eq:f-phi}  
\end{equation}%

\noindent $U_2(q)$ is also defined in \cite{RMP}.
If the correction
to the transition $2p_{1/2} - 2s_{1/2}$ is calculated in lowest order
perturbation theory using nonrelativistic point Coulomb wave functions,
the result is 205.0074\,meV, in agreement with other authors\,\cite{eides}. 

The same procedure was used to calculate the two-loop corrections; the
corresponding diagrams were first calculated by K\"allen and Sabry
\cite{kaellen}.  The Fourier transform of the corresponding potential is
given in \cite{RMP,Borie75}.  The (nonrelativistic) result for a point
nucleus is 1.5079\,meV. 

 In momentum space 
including the effect of nuclear size on the Uehling potential is
easy, since the corresponding expression for $\Delta V(q)$ is simply
multiplied by the form factor.  The numbers obtained were the same for a
dipole form factor and for a Gaussian form factor, provided the
parameters were adjusted to reproduce the assumed rms radius of the
proton. The correction can be regarded as taking into account the effect
of finite nuclear size on the virtual electron-positron pair in the
loop.  The contribution of the Uehling potential to the 2p-2s transition
is  reduced by 0.0079\,meV with a proton radius of 
0.842\,fm \cite{experiment}, and by 0.0082\,meV with a proton radius of
0.875\,fm \cite{codat02}.  
This result is consistent with the number given in \cite{eides}  
(eq.(266)).  The contribution due to the K\"allen-Sabry
potential is reduced by 0.00007\,meV.  


These numbers are modified somewhat when relativistic (Dirac) muon wave
functions are used.
The numerical values given below were calculated as the
expectation value of the Uehling potential using point-Coulomb Dirac
wave functions with reduced mass:     
\begin{center} 
    \begin{tabular}{|l|cc|cc|}
   \hline
 &                  point nucleus  &        &    $R_p$=0.875fm  &  \\
 \hline
 & $2p_{1/2}-2s_{1/2}$  & $2p_{3/2}-2s_{1/2}$  & $2p_{1/2}-2s_{1/2}$  &
 $2p_{3/2}-2s_{1/2} $           \\
Uehling          &  205.0282~  & 205.0332~ &   205.01979~ &  205.02480~  \\
Kaellen-Sabry   &    1.50814  &   1.50818  &    1.50807  &  1.50811   \\
   \hline
\end{tabular}        
\end{center}   
The effect of finite proton size calculated here was checked for several
values of the proton radius, ranging from 0.842\,fm to 0.892\,fm.  It
can be  parametrized as -0.0110$\langle r^2 \rangle$. 
The  contribution due to two and three iterations of the Uehling
potential has been calculated  by \cite{Pachuki1,karshenboim} and 
\cite{kinoshita}, respectively, giving a total of 0.1507\,meV.  This has
been verified very recently by Indelicato \cite{Indelicato}.
 An additional higher iteration including finite size and
vacuum polarization is given in ref.\,\cite{Pachuki1} (equations(66) and
(67))  and ref.\,\cite{eides} (equations(264) and (268)).  This amounts
to  -0.0164$\langle r^2 \rangle$, and is verified in Appendix~B.    
This would mean that the finite-size contributions to vacuum
polarization in muonic hydrogen can be parametrized as  
\mbox{$-\,0.0110 \langle r^2 \rangle \,-\, 0.0164 \langle r^2 \rangle$,}
giving a total of  $-0.0274\langle r^2 \rangle$.  Numerically this is  
 -0.0209(6)\,meV if the proton radius is 0.875\,fm, or -0.0194\,meV if
the proton radius is 0.842\,fm.   

Up to the time that this paper was first published the higher order effect
including nuclear size had  only been calculated for hydrogen.  
 The contribution for other light nuclei is given in
Appendix~B.  Similar results for other light atoms have recently been
presented in Ref.\cite{Karshenboim2012}.  

Corresponding numbers for muonic deuterium, calculated as the
expectation value of the Uehling potential using point-Coulomb Dirac
wave functions with reduced mass are:
\begin{center} 
    \begin{tabular}{|l|cc|cc|}
   \hline
 &                  point nucleus  &        &    $R_d$=2.130\,fm & \\
 \hline
 & $2p_{1/2}-2s_{1/2}$  & $2p_{3/2}-2s_{1/2}$  & $2p_{1/2}-2s_{1/2}$  &
 $2p_{3/2}-2s_{1/2} $           \\
Uehling          &  227.6577~  & 227.6635~ &   227.59847~ &  227.60422~  \\
Kaellen-Sabry   &     1.66622  &   1.66626  &    1.66577  &    1.66581   \\
   \hline
\end{tabular}        
\end{center}   
The effect of finite nuclear size calculated here can be  parametrized as
$-0.0130\langle r^2 \rangle$. 
However higher iterations will change these results.  
The correction for a point nucleus has been calculated for muonic
deuterium \cite{karshenboim}, giving an additional correction of
0.1718\,meV.   
The effect of finite
size as described in refs.\,\cite{eides,Pachuki1}  
is given in Appendix~B and results  in an additional contribution to the
binding energy of the 2s-state of $-0.02062\langle r^2 \rangle$, or
$-0.0936$\,meV for a deuteron radius of 2.130\,fm.  


Corresponding numbers for muonic $^3$He and $^4$He, calculated as the
expectation value of the Uehling potential using point-Coulomb Dirac
wave functions with reduced mass are:
\begin{center} 
    \begin{tabular}{|l|cc|cc|}
   \hline
 &                  point nucleus  &        &    $R_{He} \ne 0$ & \\
 \hline
 $^4$He  & $2p_{1/2}-2s_{1/2}$  & $2p_{3/2}-2s_{1/2}$  & $2p_{1/2}-2s_{1/2}$  &
 $2p_{3/2}-2s_{1/2} $           \\
Uehling          & 1666.305~   & 1666.580~ &   1665.381~ &  1665.656~  \\
Kaellen-Sabry   &     11.5731  &   11.5752  &   11.5658  &  11.5680   \\
 \hline
 $^3$He  & $2p_{1/2}-2s_{1/2}$  & $2p_{3/2}-2s_{1/2}$  & $2p_{1/2}-2s_{1/2}$  &
 $2p_{3/2}-2s_{1/2} $           \\
Uehling          & 1642.412~   & 1642.682~ &   1641.337~ &  1641.607~  \\
Kaellen-Sabry   &     11.4107  &   11.4128  &   11.4024  &  11.4045   \\
   \hline
\end{tabular}        
\end{center}   
Here the nuclear radii were taken to be   $R_{He3}$=1.966\,fm and
$R_{He4}$=1.674\,fm.  
The effect of finite nuclear size calculated here (and with other radii)
can be  parametrized as $-0.3297\langle r^2 \rangle$ for $^4$He and
$-0.3176\langle r^2 \rangle$ ($-0.3151\langle r^2 \rangle$ if the
two-loop contribution is neglected) for $^3$He.  
Higher iterations  change these results.  In \cite{RMP} they
were calculated to be about 1.70\,meV, including the effect of finite
nuclear size, for  $^4$He  and 1.4\,meV for $^3$He.  
A more recent calculation of the correction for a point nucleus has been
made for muonic $^4$He  \cite{karshenboim}, giving a correction of
1.709\,meV.  
The effect of finite nuclear size on the shift of the 2s-state is given
in Appendix~B (for both isotopes).    

The mixed muon-electron vacuum polarization correction was recalculated
and gave the same result as obtained previously, namely 0.00007\,meV.
\cite{HelvPA,eides}.  For deuterium, the contribution is 0.00008\,meV,  
for $^3$He it is 0.00200\,meV,  and for $^4$He it is 0.00208\,meV.   
For the helium isotopes, neglecting finite nuclear size would increase
the contribution by 0.0001\,meV.     

The Wichmann-Kroll \cite{WK} contribution was calculated using the
parametrization for the potential given in \cite{RMP}.  The result
obtained for hydrogen is -0.00103\,meV, consistent with that given in
\cite{eides}.  For deuterium, the contribution is  -0.00111\,meV. Values
for both helium isotopes have also been calculated.  The contribution is 
-0.01984\,meV for  $^4$He and -0.01969\,meV for  $^3$He.   
The reason for the negative sign is discussed in the review of Eides et
al. \cite{eides}.
 
The sixth order vacuum polarization corrections to the Lamb shift in
muonic hydrogen have been calculated by Kinoshita and Nio
\cite{kinoshita}.  Their result for the 2p-2s transition (in hydrogen) is 
\begin{equation*}
\Delta E^{(6)}  \,\approx \,0.00761 \, \textrm{meV} 
\end{equation*}
Recently Karshenboim et al. \cite{karshenboim,karshenboim2}
 have recalculated all of the relevant graphs,
and reproduced these results for hydrogen, including a correction that
reduced the contribution to 0.00752\,meV.  Results for muonic deuterium
(0.00842(7)\,meV) and helium  \\ 
($^3$He:\,0.073(3)\,meV and $^4$He:\,0.074(3)\,meV) have also been given.  
In addition Karshenboim et al. performed a recalculation of the  
virtual Delbr\"uck effect that is consistent with, but probably more 
accurate than, the results  given in \cite{Borie76} and \cite{RMP},
possibly due to a more accurate numerical evaluation of the sevenfold
integrals over Feynman parameters.  The published results are given as
the sum of this, the Wichmann-Kroll contribution and the previously  
uncalculated light by light contribution.  In in the case of muonic
hydrogen and helium, the virtual Delbr\"uck contribution very nearly cancels 
the Wichmann-Kroll contribution, so the total result can be expected to be 
small. 
Numerically, they found the following total "light-by-light"
corrections (i.e. the sum of the Wichmann-Kroll, virtual Delbr\"uck and
previously uncalculated term), in meV for all cases of interest:
\begin{center} 
    \begin{tabular}{|lr|}
   \hline
$\Delta\,E(\mu H)$          & -0.00089(2)       \\
$\Delta\,E(\mu D)$          & -0.00096(2)       \\
$\Delta\,E(\mu ^3He)$       &  -0.0134(6)       \\
$\Delta\,E(\mu ^4He)$       &  -0.0136(6)       \\
   \hline
\end{tabular}        
\end{center}

The hadronic vacuum polarization contribution has been estimated by a
number of authors \cite{hadron,friar99,RMP,eides}.  It amounts to about
0.012\,meV in hydrogen, 0.013\,meV in deuterium, 0.219\,meV in $^3$He
and 0.225\,meV in $^4$He.  An uncertainty of about 5\% should be expected.


\subsubsection*{Finite nuclear size and nuclear polarization }
\vspace{-0.2cm}
The main contribution due to finite nuclear size has been given
analytically to order $(\alpha Z)^6$ by Friar \cite{friar79}.  The main
result is 
\begin{equation}
\Delta E_{ns}\,=\,-\frac{2 \alpha Z}{3} \left(\frac{\alpha Z m_r}{n}\right)^3
 \cdot \left[\langle r^2 \rangle - \frac{\alpha Z m_r}{2} \langle r^3
 \rangle_{(2)} +(\alpha Z)^2 (F_{REL}+m^2_r F_{NR})  \right]
\label{eq:FS-friar}
\end{equation}
Radiative corrections to the main term have been calculated by Eides and
Grotch \cite{eides-grotch}.  They contribute an additional correction of 
\begin{equation*}
\Delta E_{ns}\,=\,-\frac{2 \alpha Z}{3} \left(\frac{\alpha Z m_r}{n}\right)^3
 \cdot \alpha^2 Z (23/4 - 4 \ln(2) -3/4)  \cdot \langle r^2 \rangle
\end{equation*}
This correction (of order $\alpha (\alpha Z)^5$) amounts to an
additional (fractional) correction to the main term equal to 
$2.2275 \alpha^2 Z \langle r^2 \rangle \,=\,1.1862 \times 10^{-4} Z \langle r^2 \rangle $.   
For muonic hydrogen, the main coefficient of $\langle r^2 \rangle$ is
\mbox{-5.1973\,meV fm$^{-2}$} without the radiative correction; it is
modified  to \mbox{-5.1979\,meV fm$^{-2}$} with the correction.  
For the main term, the shift is
-3.979$\pm$0.076\,meV if the proton rms 
radius is 0.875$\pm$0.007)\,fm and -3.685$\pm$0.008\,meV  
if the proton rms radius is 0.842$\pm$0.001)\,fm. The radiative
correction increases these numbers by 0.0005\,meV.   
Assuming a proton rms radius of 0.875\,fm, the
correction from the first two terms in Eq.(\ref{eq:FS-friar}) to the
2s-level is 3.956\,meV for a dipole form factor and  
3.958\,meV for a Gaussian form factor.  The parameters were fitted to
the proton rms radius.  The difference is due to the second term in   
Eq.(\ref{eq:FS-friar}).  It contributes 
0.009126 $\langle r^3 \rangle_{(2)}$\,meV (or 0.0232\,meV for a dipole form
factor and 0.0212\,meV for a Gaussian form factor).    

 Pachucki \cite{Pachuki2} has estimated a correction similar to 
the second term (proportional to $\langle r^3 \rangle_{(2)} $) in
Eq.(\ref{eq:FS-friar}).  Friar and Sick \cite{friar04} have shown that
the results are equivalent, at least for static potentials.  
\newpage
Since the second term in  Eq.(\ref{eq:FS-friar}) is numerically
important, it has been investigated further, at least for the case of
hydrogen.  The third Zemach moment was calculated in a model-independent
manner from electron-proton scattering data by Friar and Sick
\cite{friar04}, with the result  
$\langle r^3 \rangle_{(2)}$\,=\,2.71(13)fm$^{3}$, for a contribution of
0.0247\,meV, with an uncertainty of 0.0012\,meV.  
and more recently, using new  experimental results, by
Distler~et~al.\cite{newzemach}, with the result  
$\langle r^3 \rangle_{(2)}$\,=\,2.85(8)fm$^{3}$, for a contribution of
0.0260\,meV, with an uncertainty of 0.0007\,meV.   The relationship
between the third Zemach moment and the charge radius will be discussed
below. 

Since the coefficient of $\langle r^2 \rangle$ is important for the
determination of the nuclear radius, the terms of order $(\alpha Z)^6$
in  Eq.(\ref{eq:FS-friar}) that depend only on this quantity are
separated from the other terms.   This affects only $(\alpha Z)^2 F_{REL}$. 
Details will be given in Appendix~B.  The result is an extra
contribution to the energy shift of the 2s-state equal to 
\begin{equation*}
\Delta E_{2s}\,=\,\frac{2}{3}\frac{(\alpha Z)^6 m_r^3}{n^3} \langle r^2 \rangle 
 [\gamma - \frac{35}{16} + \ln(\alpha Z)] 
\end{equation*}
($\gamma$ is Euler's constant). This results in an energy shift of 
$-0.00181\langle r^2 \rangle\,=\,-0.00138$\,meV in hydrogen.  The
value of the coefficient is compatible with the estimated contribution
of $-0.0016\langle r^2 \rangle$ given in ref.\,\cite{experiment}.
The other 
remaining terms (of order $(\alpha Z)^6$) given in \cite{friar79} 
contribute 0.000123\,meV if an exponential charge distribution is 
used. 
 Some details are given in Appendix~B.  This estimate includes all 
of the terms in Eq.(\,\ref{eq:FS-friar}), while other authors
\cite{eides,Pachuki2} give only some of them.   

Friar \cite{friar79} also found finite size contributions 
to the binding energy of the 2p-levels, , amounting to 
\begin{subequations}
\begin{gather}
\Delta E_{np1/2}\,=\,\frac{(\alpha Z)^3}{3}\left(\frac{\alpha Z m_r}{n}\right)^3
 \cdot (1-1/n^2)\left[\frac{\langle r^2 \rangle}{2} + \frac{m_r^2 \langle r^4
 \rangle}{15}   \right]  \\
\Delta E_{np3/2}\,=\, (\alpha Z) \left(\frac{\alpha Z m_r}{n}\right)^5
 \cdot \frac{(n^2-1)\langle r^4 \rangle}{45}
\end{gather}
\label{eq:FS-P}
\end{subequations}
If terms of order $(\alpha Z)^6$ are included in the coefficient of 
$\langle r^2 \rangle$ are to be taken into account consistently, the
contribution to the 2p$_{1/2}$ energy level must also be included.  For
hydrogen, this is -0.0000519$\langle r^2 \rangle$\,=\,-4$\cdot10^{-5}$\,meV. 
Subtracting this contribution from the order $(\alpha Z)^6$ shift of the
2s level ( $b_c \langle r^2 \rangle $, see Appendix\,B)   results in a
 contribution to the 2s$_{1/2}$-2p$_{1/2}$ transition of 0.00134\,meV,
 in good agreement with previous estimates \cite{Pachuki2}.   
There is no contribution to the coefficient of $\langle r^2 \rangle$ for
transitions to the 2p$_{3/2}$ state. The fine structure is, of course, affected.
The contribution to the 2p$_{1/2}$ energy level of  deuterium gives  a
correction of -0.0000606$\langle r^2 \rangle$\,=\,-2.8$\cdot10^{-4}$\,meV.
The contribution to the energy of the 2p$_{3/2}$~state is much smaller, 
 and is proportional to $\langle r^4 \rangle$;  the 
relation to the rms-radius of the nucleus is model dependent.  Analytic
expressions for  $\langle r^4 \rangle$ in terms of $\langle r^2 \rangle$
can be derived from the table given in ref.\,\cite{friar79}.  For an
exponential charge distribution, 
$\langle r^4 \rangle\,=\,5(\langle r^2 \rangle)^2/2$ and for a Gaussian
charge distribution $\langle r^4 \rangle\,=\,5(\langle r^2 \rangle)^2/3$.
For the helium isotopes these corrections are significant for both levels.

These contributions, with numerical values for the coefficients and
energy shifts 
for deuterium and helium will be given in Appendix~B. 

For the extraction of a single radius parameter from experimental data,
it is useful, and has become customary, to write the Lamb shift in the
form  
\begin{equation*}
\Delta E_{LS}\,=\, A + B \langle r^2 \rangle + C (\langle r^2 \rangle)^{3/2}
\end{equation*}
 It is straightforward to determine $B$.  There are several
 contributions, and these will be  given in Appendix~B, with numerical
 values for all cases.  
However, the value of $C$ depends
 on a model for the charge distribution, since it depends on the ratio 
\begin{equation*}
 f\,=\, \langle r^3 \rangle_{(2)}/( \langle r^2 \rangle)^{3/2}
\end{equation*}
For Gaussian and exponential charge distributions, $f$ can be calculated
analytically  \cite{friar79}, with the result 
$f=32/(3\sqrt{3 \pi})$\,=\,3.4745 for a Gaussian charge distribution,  
and \\ $f=105/(16\sqrt{3})$\,=\,3.7889 for an exponential charge
distribution, which corresponds to a dipole form factor.  The
corresponding coefficients of $( \langle r^2 \rangle)^{3/2}$ (in units
meV\,fm$^{-3}$) are, for hydrogen, 
0.03458~(dipole form factor) and 0.03171~(Gaussian form factor).
When the third Zemach radius is determined from experiment, the value of
$f$ must also be determined from the same experimental data.  Using the
results of Friar and Sick~\cite{friar04,sick} gives $f$\,=\,3.780(52),
while the newest evaluation~\cite{newzemach} gives  $f$\,=\,4.18(13),
The corresponding coefficients of  $(\langle r^2 \rangle)^{3/2}$ are 
0.0345(5) and 0.0383(12), respectively.  In the newest CODATA
compilation \cite{codat06} a value of $f=3.4(2)$ 
is given.  
This is almost certainly too small in view of the previous values
discussed.  An average value of the two model independent determinations
corresponds to  $f=4.0(2)$  and would result in a coefficient of 
0.0365(18)\,meV\,fm$^{-3}.$  
The uncertainty in the precise value of the
coefficient  will have consequences for the theoretical uncertainty in
the value of the proton radius, evaluated as in the recent experiment
\cite{experiment}.  The value of $A$ also has a theoretical uncertainty,
which may have been slightly underestimated. 
(see the summary tables and Appendix~C).

The recommended contribution due to the third Zemach moment (expressed
in terms of $(\langle r^2 \rangle)^{3/2}$) for deuterium is
(0.0112$f$\,=\,0.0448(22)): 
0.0448(22)\,meV\,fm$^{-3} (\langle r^2 \rangle)^{3/2}.$

For the helium isotopes a Gaussian charge distribution is a fairly good
approximation \cite{friar79}, so a value of  use $f=3.5(.1)$  should be
a sufficiently good approximation.  Then 
C\,=\,1.40(4)\,meV\,fm$^{-3} (\langle r^2 \rangle)^{3/2}$ for  $^4$He and
1.35(4)\,meV\,fm$^{-3} (\langle r^2 \rangle)^{3/2}$ for  $^3$He.

The contribution due to nuclear polarization (in hydrogen) has been
calculated by  Rosenfelder \cite{rosenfelder99} to be
0.017\,$\pm$\,0.004\,meV, and by Pachucki \cite{Pachuki2} to be
0.012\,$\pm$\,0.002\,meV.  Other calculations \cite{srartsev,faustov}
give intermediate values (0.013\,meV and 0.016\,meV, respectively).  
A very recent calculation \cite{carlson11} gives a unified treatment of
elastic and inelastic contributions.  The total given there is very
close to the total that would be obtained in this work if only the
inelastic contribution (0.0127(5)) is included in the nuclear
polarization correction, and the Zemach moment contribution is
interpreted as the sum of the other two contributions given there.
However, the estimated error in the polarizability correction has been
enlarged to be comparable to the total uncertainty.

The contribution due to nuclear polarization in deuterium has been
calculated by Leidemann and Rosenfelder \cite{rosenfelder95} to be 
1.50\,$\pm$\,0.025\,meV (see also \cite{rosenfelder93}). These
calculations may have neglected possible "elastic" contributions that
would be comparable to the quoted uncertainty.  A newer
calculation \cite{pachucki-pol} gives 1.680\,$\pm$\,0.016\,meV.  The
author also claims that some contributions to the
polarizability cancel the contribution due to the Zemach moment.  This
has been verified by Friar (\cite{friar13}). He obtained 1.942\,meV
for this correction.  Other recent calculations (\cite{ji,carlson14}) 
give 1.698\,$\pm$\,0.20\,meV and 2.01\,$\pm$\,0.74\,meV, respectively. 
Although a possible contribution from the Zemach moment is given here 
(it would be 0.433(21)\,meV), it is not included in subsequent summaries. 
In any case, it is clear that an explanation of the differences in 
these calculated results is needed.  A part of the difference between the 
numbers given in references \cite{pachucki-pol,friar13} and \cite{ji} may 
be due to higher order terms which one or the other neglected. 

An estimate of the nuclear polarization contribution for the helium
isotopes is given in \cite{RMP,friar77}.  It amounts to 3.1\,$\pm$\,0.6\,meV 
for $^4$He and  4.9\,$\pm$\,1.0\,meV for $^3$He.  A newer calculation
for $^4$He \cite{ji} gives a value of  2.47\,$\pm$\,0.15\,meV, as well
as a contribution for the Zemach moment of 6.32\,meV. 

\subsubsection*{Relativistic Recoil }
\vspace{-0.2cm}
As is well-known, the center-of-mass motion can be separated exactly
from the relative motion only in the nonrelativistic limit.
Relativistic corrections have been studied by many authors, and will not
be reviewed here.  The relativistic recoil corrections summarized in
\cite{RMP} include the effect of finite nuclear size to leading order in
$m_{\mu}/m_N$ properly.  In fact the effect calculated here is probably
accurate to order $\alpha Z m_{\mu}/m_N$ 

Up to 2004 this method was not used to treat recoil corrections to
vacuum polarization.  The recoil correction to the energy shift due to 
 the Uehling potential can be included explicitly, 
as a perturbation correction to point-Coulomb values, in a  manner 
very similar to that  described in \cite{RMP} (and in \cite{friar79}).
This was first described in \cite{Borie05};  the result in the case of
hydrogen  for the total relativistic correction agrees very well with an
independent calculation based on a 
generalized Breit equation  \cite{Pachuki3}.  The basic formulas and
numerical results are given here.  Details are given in  Appendix~A.  
To leading order in $1/m_N$, the energy levels including a shift beyond
that due to using the reduced mass in the Dirac equation  are given by 
\begin{equation}
  E~=~E_r - \frac{B_0^2}{2  m_N} + \frac{1}{2 m_N} \langle h(r) +  
  2 B_0 P_1(r) \rangle
\label{eq:rmp1}
\end{equation}
where $E_r$ is the energy level calculated using the Dirac equation with
reduced mass and  $B_0$ is the unperturbed binding energy.  Also
\begin{equation}
   h(r)\, = \, - P_1(r)(P_1(r) + \frac{1}{r} Q_2(r))  
            - \frac{1}{3 r} Q_2(r) [P_1(r) + \frac{1}{r^3} Q_4(r)]
\label{eq:rmp2}
\end{equation}
where   $P_1,\,Q_2,$\,and $Q_4$ are defined in references \cite{RMP} and
\cite{Borie05} and in Appendix A (Eq.(\ref{eq:rmp3})). 
Details of the calculation for the case of vacuum polarization are given 
in Appendix\,A and in Ref.\cite{Borie05}.  
Quite recently it has been pointed out that this
calculation may have been done in a gauge that missed some contributions 
to retardation or two-photon-exchange effects \cite{Karshenboim2012}.    
\begin{table}[!h]
\begin{center} 
  \begin{tabular}{|lrrr|}
 \hline
  &          Main term         &   Correction  &   Total  \\
     \hline
hydrogen     &  0.02084    &   -0.00208    &   0.01876    \\
deuterium    &  0.02271   &   -0.00093     &   0.02178    \\
$^3$He       & 0.5233~     &    -0.0140~   & 0.50934        \\
$^4$He       & 0.5301~  &    -0.0090~   & 0.52110        \\
 \hline
\end{tabular}        
 \caption{ Total relativistic corrections due to vacuum polarization  to
 the Lamb shift in light muonic 
 atoms (in meV). The main term is the difference between the vacuum
 polarization contribution  calculated with relativistic and
 nonrelativistic wave functions.  }
  \label{tab:RR-recoil}
\end{center}   
\end{table}    
To obtain the full relativistic and recoil corrections,
one must add the difference between the expectation values of the
Uehling potential calculated with relativistic and nonrelativistic
wave functions.  
The results for all cases are summarized in Table\,\ref{tab:RR-recoil} 
These results are based on corrected values  given in 
Ref.\,\cite{Karshenboim2012}.    
The treatment presented here has the
advantage of treating the main contribution relativistically and applying
a small correction that can be calculated using first order perturbation
theory.  
The purely relativistic correction to the K\"allen-Sabry contribution
(i.e. to the main term) was not included.   
For hydrogen and deuterium, this contributes approximately 0.0002\,meV to 
the main term, For the  helium isotopes the contribution is approximately 
0.0037\,meV.  These are so small that relativistic recoil 
corrections may be superfluous.    


This method can also be used to calculate relativistic recoil corrections 
to the shift due to finite nuclear size  when a model for the charge  
distribution is given.  This was done by Friar \cite{friar79}; 
The recoil correction is given by: 
\begin{equation*}
  \Delta E_r~=~- \frac{(\alpha Z)^2}{m_N} \big(\frac{\alpha Z m_r}{n}\big)^3
   \langle r \rangle_{(2)}   \delta_{\ell 0}.
\end{equation*}
Numerically the contribution due to finite nuclear size to the recoil
correction for 
the binding energy of the 2s-level  is -0.013\,meV in hydrogen, and
 -0.019\,meV in deuterium.  The factor $1/m_N$
 is replaced by  $1/(m_{\mu}+m_N)$, also consistent with the
 calculations presented in \cite{friar79}.  In the case of hydrogen, the
 contribution is approximately 0.004\,meV smaller if a Gaussian charge
 distribution is used instead of an exponential charge
 distribution. With the exponential charge distribution, the value of
 the contribution ranges from 0.0131\,meV to 0.0139\,meV as the charge
 radius varies from 0.842\,fm to 0.895\,fm.   A calculation with the
 value given by Distler et al. \cite{newzemach} is also in this range.
This contribution is thus  nearly constant for hydrogen.  
For muonic $^4$He the recoil contribution is -0.267\,meV for a
Gaussian charge distribution and an rms-radius of 1.681\,fm.  Varying
the radius by $\pm$\,0.02\,fm changes the contribution by
$\pm$0.003\,meV.  
For muonic $^3$He this recoil contribution is -0.404\,meV for a
Gaussian charge distribution and an rms-radius of 1.966\,fm.  
There has been some discussion as to whether or not this contribution is 
already taken into account in calculations of two-photon-exchange effects.

The review by Eides et.al \cite{eides} gives a better version of the two
photon recoil (Eq. 136) than was available for the review by Borie and
G. Rinker \cite{RMP}.  Evaluating this expression for muonic hydrogen
gives a contribution of -0.04497\,meV to the 2p-2s transition in
hydrogen,  -0.02656\,meV in deuterium,  -0.4330\,meV in $^4$He, and 
 -0.5581\,meV in $^3$He. 

Higher order radiative recoil corrections give  additional
contributions \cite{eides}.   These correspond to (their) Table~9 and two
additional terms in Table~8.  Numerically, the contributions are 
-0.01003\,meV for hydrogen, -0.00302\,meV for deuterium, -0.04737\,meV 
for $^4$He and   -0.08102\,meV for $^3$He. 

An additional recoil correction for states with $\ell \, \ne \, 0$ has
been given by \cite{barker} (see also \cite{eides}).  It is 
\begin{equation}
\Delta E_{n, \ell ,j}~=~\frac{(\alpha Z)^4 \cdot m_r^3}{2 n^3 m_N^2} 
 (1-\delta_{\ell 0}) \left(\frac{1}{j+1/2} - \frac{1}{\ell+1/2}\right)
\label{eq:recoil1}
\end{equation}
Note that $1/(j+1/2)-1/(\ell+1/2) = 1/(\kappa (2\ell+1))$.  
When evaluated for the 2p-states of muonic hydrogen, one finds a
contribution to the 2p-2s transition energy of 0.05748\,meV for the
p$_{1/2}$ state and -0.02873\,meV for the p$_{3/2}$ state.  For $^3$He
the contribution is  0.12654\,meV for the p$_{1/2}$ state and
-0.06327\,meV for the p$_{3/2}$ state.  
For integer spin nuclei, there is an additional shift of the 2s-state,
amounting to $-\alpha Z(\alpha Z m_r/n)^3/(2m_N^2)$ \cite{kp-sk}. This
will not affect the fine structure. 
Numerical values for the energy shift of the 2p$_{1/2}$ and 2p$_{3/2}$ states of
deuterium are: 0.01681\,meV and -0.00840\,meV, respectively.  The total
contribution for the 2p$_{1/2}$-2s transition is then 0.06724\,meV.  Numerical 
values for  $^4$He are 0.07379\,meV for the 2p$_{1/2}$ state, -0.03690\,meV 
for the 2p$_{3/2}$ state, and -0.22137\,meV for the 2s state.   

A final point about recoil corrections is that in the case of light
muonic atoms, the mass ratio $m_{\mu}/m_N$ is considerably larger than
the usual perturbation expansion parameter $\alpha Z$.  

\subsubsection*{Muon Lamb Shift }
\vspace{-0.2cm}
For the calculation of muon self-energy and vacuum polarization, the
lowest order (one-loop approximation) contribution is well-known, at
least in perturbation theory.  For a point nucleus, and neglecting
possible contributions due to vacuum polarization, so that 
$\nabla^2 V = -4 \pi \alpha Z \rho$ ($\rho$ is the nuclear charge
density) is approximated by  a delta function, giving $V\,=\,- \alpha Z/r$ and 
\begin{equation*}
\langle\nabla^2 V\rangle\,=\,-4\pi \alpha Z |\psi_{ns}(0)|^2 \delta_{\ell 0}
\,=\,-4 Z \alpha \big(\frac{Z \alpha m_r}{n}\big)^3 \delta_{\ell 0}
\end{equation*}
 one finds a  contribution of
-0.66788\,meV for the \mbox{$2s_{1/2}-2p_{1/2}$} transition and 
 -0.65031\,meV for the  \mbox{$2s_{1/2}-2p_{3/2}$} transition.  
For deuterium, the corresponding contributions are given by 
-0.77462\,meV for the \mbox{$2s_{1/2}-2p_{1/2}$} transition and 
 -0.75512\,meV for the  \mbox{$2s_{1/2}-2p_{3/2}$} transition.  
These numbers include  muon vacuum polarization
(0.016844\,meV for hydrogen,  0.019682\,meV in deuterium,
 0.33225\,meV in $^3$He, and  0.34132\,meV in $^4$He), 
and an extra term of order $(Z \alpha)^5$  as given in \cite{eides} for
the 2s-state:
\begin{equation*}
\Delta E_{2s}\,=\, \frac{\alpha (\alpha Z)^5 m_{\mu} }{4} \cdot
\left(\frac{m_r}{m_{\mu}}\right)^3   \cdot    
 \left(\frac{139}{64} + \frac{5}{96} - \ln(2) \right)
\end{equation*}
which contributes -0.00443\,meV for hydrogen and  -0.00518\,meV for deuterium.
For the helium isotopes the contributions are  -0.17490\,meV in $^3$He
and -0.17967\,meV  in $^4$He.  
\begin{table}[!h]
\begin{center} 
    \begin{tabular}{|lrr|}
      \hline
Transition   &  $2p_{1/2}-2s_{1/2}$  & $2p_{3/2}-2s_{1/2}$            \\
\hline
Hydrogen   &       &     \\           
second order          &  -0.667882   &    -0.650313            \\
higher orders        &   -0.001714    &   -0.001647        \\
         &      &     \\   
Total     &     -0.669596   &    -0.651960   \\
\hline
Deuterium   &       &     \\           
second order          &   -0.774616 &    -0.755125            \\
higher orders        &   -0.002001     &   -0.001926        \\
      &         &          \\
Total     &     -0.776617   &    -0.757051   \\
\hline
$^3$He   &       &     \\           
second order          &   -10.827368 &    -10.504167           \\
higher orders        &     -0.033749     & -0.032515        \\
      &         &          \\
Total     &     -10.861117   &    -10.536672   \\
\hline
$^4$He   &       &     \\           
second order          &   -11.105708 &    -10.776650           \\
higher orders        &     -0.034663     &  -0.033406        \\
      &         &          \\
Total     &     -11.140370   &    -10.810056   \\
\hline
\end{tabular}        
 \caption{ Contributions to the muon Lamb shift  
($ E(2p_{1/2}) - E(2s_{1/2})$)  in muonic  hydrogen, deuterium, $^3$He,
and $^4$He,  in meV.   }
  \label{tab:muonLS}
\end{center}   
\end{table} 
These results, and the higher order corrections \cite{RMP,HelvPA} 
are summarized in Table\,\ref{tab:muonLS}. 

The higher order contributions can be written in the form
\begin{equation*}
\Delta E^{4,6}_{LS}\,=\,\frac{1}{m_{\mu}^2} \cdot \langle\nabla^2 V\rangle
 \big[m_{\mu}^2 F'_1(0) + \frac{a_{\mu}}{2} \big] + \frac{a_{\mu}}{m_{\mu} m_r}
\Big\langle \frac{1}{r}\,\frac{dV}{dr} \vec{L} \cdot \vec{\sigma}_{\mu}
 \Big\rangle 
\end{equation*}
where $F_2(0) \,=\,a_{\mu}$; the higher order contributions (fourth and
sixth) can be taken from the well-known theory of the muon's anomalous
magnetic moment: \\ \mbox{$F_2(0) \,=\,a_{\mu}=\alpha/2 \pi +
  0.7658(\alpha/\pi)^2 + 24.05(\alpha/\pi)^3$}.   
The fourth order contribution to $F'_1(0)$ is 
\mbox{$0.46994(\alpha/\pi)^2 + 2.21656(\alpha/\pi)^2
  \,=\,2.68650(\alpha/\pi)^2$} \cite{RMP,HelvPA}.  
These expressions include the  fourth order electron loops
\cite{barbieri73} which dominate the  
 fourth order contribution. The contribution of the electron loops alone 
is  -0.00168\,meV for the $2s_{1/2}-2p_{1/2}$ transition and  -0.00159\,meV
for the $2s_{1/2}-2p_{3/2}$ transition.  Including the rest, which is
the same as for the electron \cite{RMP}, gives -0.00169\,meV and 
-0.00164\,meV, respectively.  Since  it is no more
difficult to include the complete contribution to $F'_1(0)$ and
$F_2(0)$ in the analysis, there is no reason not to do so.  An
additional contribution due to including the K\"allen-Sabry correction
with muon loops effectively adds $-(41/162)(\alpha/\pi)^2$ to  $F'_1(0)$
\cite{eides}.  This has been included in the higher order corrections.

 The numbers given in
Table\,\ref{tab:muonLS} were calculated assuming that $\nabla^2 V$ can
be approximated by a delta function.
 However, the potential should be corrected for vacuum polarization due
 to electron loops, and, at least for s-states, for the effect of finite
 nuclear size.  This has been done a long time ago
 (\cite{hyperfine,Brodsky,sternheim,zemach})  for the hyperfine
 structure,  but up to now not completely for the  "muon Lamb shift". 
In addition, a correction as a result of  
distortion of the wave function of the 2s-state due to vacuum
polarization,  has recently been 
calculated by Ivanov et al. \cite{ivanov}.  Numerically the results
agree with previous calculations \cite{Borie05,Borie-d05} for the case
of the hyperfine structure of the 2s-state (as given in
Eq.\,\ref{eq:epsvp2}), when the effect of finite nuclear size is neglected.   

Pachucki \cite{Pachuki1} has estimated an additional contribution of
-0.005\,meV  corresponding to a vacuum polarization
insert in the external photon for hydrogen.  
It is not clear to what extent this contribution is described by corrections 
to $\langle\nabla^2 V\rangle$ as described in Appendix~C.   A more
recent calculation by Jentschura \cite{jentschura} gives a different
result for this contribution.  An independent recalculation is desirable
and this will be presented for a part of the contribution in Appendix~C.
The effect of finite nuclear size (analogous to the Bohr-Weisskopf
effect known for the hyperfine structure) will also be estimated there.

The numerical value of the partial correction to the 2s-2p transition
calculated there is of the order of 0.002\,meV for hydrogen, which is
too small to account for the discrepancy in the proton radius.  It is 
not included in Table\,\ref{tab:muonLS}, since corrections to the Bethe
sum have not been calculated.  However, this introduces  an uncertainty
in the theoretical value.  Also the corrections are larger for other
light nuclei.  The total corrections calculated in Appendix~C should be
considered as  uncertainties in the theoretical value for the Lamb shift.  

\subsubsection*{Summary of contributions }
\vspace{-0.2cm}
Using the fundamental constants from the CODATA 2006 compilation
(\cite{codat06}),   except where noted,  
one finds the  transition energies for hydrogen in meV in Table \ref{tab:final}. 
Here the main  vacuum polarization contributions are given for a   
point nucleus, using the Dirac equation with reduced mass. Relativistic
recoil corrections are given separately.  
  Some uncertainties have been increased from the values
given by the authors, as discussed in the text. 
\begin{table}[!h]
\begin{center} 
    \begin{tabular}{|lrr|}
      \hline
Contribution  &  Value (meV) &  Uncertainty (meV) \\
      \hline
Uehling        &     205.0282~    &  \\
K\"allen-Sabry     &       1.5081~  &           \\
VP iterations \cite{Pachuki1,karshenboim}  & 0.1507~  &    \\
sixth order \cite{karshenboim}  & 0.00752    &   \\
Total "LBL"  \cite{karshenboim2} &  -0.00089 &  0.00002  \\ 
mixed mu-e VP    &       0.00007   &  \\
hadronic VP       &      0.011~~~    &   0.001~~~     \\       
\hline
recoil  \cite{eides} (eq136)     & -0.04497  &  \\
recoil, higher order \cite{eides} & -0.0100~ & \\
recoil, finite size \cite{friar79} & 0.013~~  &  0.001~~~ \\
recoil correction  to VP \cite{Karshenboim2012}  &  -0.0021~ &  \\
additional recoil  \cite{barker} & 0.0575~ &   \\
\hline
muon Lamb shift   &    &  \\
second order      &     -0.66788 & \\
higher orders      &     -0.00171 &  \\
\hline
nuclear size  ($R_p$=0.875\,fm)  & &  0.007\,fm \\
main correction $B\cdot\langle r^2 \rangle$  & -4.002~~    &  0.064~~~   \\
Zemach moment \cite{friar79} &  0.0244~  &  0.002~~~ \\
remaining order $(\alpha Z)^6$ \cite{friar79} &  -0.0001~  &  \\
polarization      &    0.0127~ &  0.003~~~          \\
correction to the 2p$_{1/2}$ level   &   0.00004  &  \\
%
\hline
\end{tabular}        
 \caption{ Contributions to the muonic hydrogen Lamb shift. The proton
   radius is taken from \cite{codat02}. }
  \label{tab:final}
\end{center}   
\end{table}   
\vspace{-0.1cm}     
\begin{center}
 \begin{minipage}{\linewidth}
In the case of the muon Lamb shift, the numbers in Table\,\ref{tab:final}
are for the $2s_{1/2}-2p_{1/2}$ transition.   
\end{minipage}
\end{center}   
\vspace{-0.1cm}     
As calculated in appendix~B (see Table\,\ref{tab:coeff}), the finite
size corrections for hydrogen can be parametrized as 
\mbox{$- 5.22718 \langle r^2 \rangle \,+\,0.00913 \langle r^3 \rangle_{(2)}$,}  
where energies are in meV and radii in fm.   
The total coefficient of $\langle r^2 \rangle$ differs slightly from
that given in 
ref.\,\cite{experiment}  \mbox{(-5.2262meV\,fm$^{-2}$).} The difference is 
due partly to a more precise determination of the coefficient of order
$(\alpha Z)^6$  and partly to the inclusion of the radiative correction 
given  by Eides and Grotch \cite{eides-grotch}. Note that the "remaining
order ($(\alpha Z)^6$)" corrections refer to contributions not proportional 
to $\langle r^2 \rangle $ given in Appendix\,B and in Eq.(\ref{eq:FS-friar}).  
They are included in the total given below.  For hydrogen the
contribution is very small. The shift  of the $2p_{1/2}$-state is
included in the table, but it is given separately only for
informational purposes and is not counted a second time in sums.  This
holds for all other cases considered.  The second term can be approximated by 
$0.0365(18) (\langle r^2 \rangle)^{3/2}$.  This results in a total transition 
energy (without hyperfine interaction) of  \\
\mbox{$206.0484(60)+0.0127-5.22718 \langle r^2 \rangle\,+\,0.0365(18)\langle r^2 \rangle^{3/2}$}.  


If the proton
radius is taken to be 0.842(1)\,fm,  the (total) nuclear size correction
becomes $-3.6855\pm 0.010$\,meV.  With a radius of 0.875(7)\,fm, it is
$-3.978\pm 0.065$\,meV.  

A recent paper by Carroll et al. \cite{carroll} calculates the Lamb
shift in muonic hydrogen relativistically and nonperturbatively.  This
is  very useful progress in removing the limitations of perturbation
theory.   For what is supposed to be the same calculated transition
energy they obtain 
\mbox{$206.0604-5.2794 \langle r^2 \rangle\,+\,0.0546\langle r^2 \rangle^{3/2}$.}  
It would have been helpful if they had provided more details about how
they obtain corrections due to two- and three-loop vacuum polarization
contributions (Kaellen-Sabry, sixth order, and higher orders), muon
self energy and muon vacuum polarization, hadron vacuum polarization,
relativistic recoil, and nuclear polarization.  Instead they simply used
the values listed in the supplement to ref.\,\cite{experiment}. 
Also, there are some problems with their calculation of the hyperfine
splitting of the 2s state. 
A very recent nonperturbative calculation by Indelicato \cite{Indelicato} 
verified many of these contributions.

\subsubsection*{Summary of contributions for muonic deuterium }
\vspace{-0.2cm}
Muonic deuterium is in many ways similar to muonic hydrogen, but there
are some differences.  In addition to the different mass,  
the deuteron has spin 1 and both magnetic and quadrupole moments.  
\begin{table}[!h]
\begin{center} 
    \begin{tabular}{|lrr|}
      \hline
Contribution  &  Value (meV) &  Uncertainty (meV) \\
      \hline
Uehling        &     227.6577~    &  \\
K\"allen-Sabry     &       1.6662~  &           \\
VP iterations \cite{karshenboim}  &  0.1718~  &    \\
sixth order \cite{karshenboim}  & 0.00842    & 0.00007~  \\
Total "LBL"  \cite{karshenboim2} &  -0.00096 &  0.00002~  \\ 
mixed mu-e VP    &       0.00008   &  \\
hadronic VP       &      0.013~~~    &   0.001~~~     \\       
\hline
recoil  \cite{eides} (eq136)     & -0.02656  &  \\
recoil, higher order \cite{eides} & -0.00302 & \\
recoil, finite size \cite{friar79} & 0.019~~~  &  0.003~~~ \\
recoil correction  to VP \cite{Karshenboim2012}  &  -0.00093 &  \\
additional recoil  \cite{barker} & 0.06724 &   \\
\hline
muon Lamb shift   &    &  \\
second order      &     -0.774616 &    \\
higher orders      &     -0.002001 &  \\
\hline
nuclear size  ($R_d$=2.130\,fm \cite{sick98})  & &  0.003\,fm \\
main correction $B\cdot\langle r^2 \rangle$  & -27.718~~    &  0.078~~~   \\
remaining order $(\alpha Z)^6$ \cite{friar79} &  0.0033~  &  \\
polarization                  &    1.690~~ &  ~~0.06~~~~          \\
correction to the 2p$_{1/2}$ level   &   0.00038  &  \\
\hline
\end{tabular}        
 \caption{ Contributions to the muonic deuterium Lamb shift. The 
 deuteron  radius is taken from \cite{sick98}. }
  \label{tab:Total-D}
\end{center}   
\vspace{-0.15cm}    
\end{table}        
For deuterium, 
one finds the  transition energies in meV in Table\,\ref{tab:Total-D}.  
Also here the main vacuum polarization contributions are given for a   
point nucleus, using the Dirac equation with reduced mass.  
The nuclear polarization (plus Zemach moment contribution) correction is
taken to be a guess based on the most recent results
\cite{pachucki-pol,friar13,ji,carlson14} given previously.  As before,
the energy shift of the 2p$_{1/2}$ level is given separately only for
informational purposes and is not counted a second time in sums. 

As calculated in appendix~B (see Table\,\ref{tab:coeff}), the finite
size corrections for the 2s-2p transition in muonic deuterium 
can be parametrized as 
$-6.10940 \langle r^2 \rangle $
where energies are in meV and radii in fm.   For more
details, including the contribution for remaining order $(\alpha Z)^6$,
see appendix~B.   
The total transition energy (without hyperfine structure) calculated
here is then 
\begin{equation*}
(228.7797\,\pm\,0.04\,+\,1.690\,\pm\,0.06\,-\,6.10940 \langle r^2 \rangle)\,meV 
\end{equation*}%

\newpage
A very recent calculation of the same transition energy by Krutov and
Martynenko \cite{martynenko11}  gives similar results, with one significant
difference. For the nuclear structure contributions of order $(Z\alpha)^5$ 
they use the value given by Pachucki\,\cite{pachucki-pol}, which
supposedly includes nuclear polarization and the Zemach moment term. 
In this work, the more standard separation has been presented; however, the
combined polarization is taken as guess based on the most recent 
calculations \cite{pachucki-pol,friar13,ji,carlson14}.


\subsubsection*{Summary of contributions for muonic helium }
\vspace{-0.2cm}
For $^4$He   
one finds the  transition energies in meV in table \ref{tab:Total-He4}.  
Also here the main  vacuum polarization contributions are given for a   
point nucleus, using the Dirac equation with reduced mass.  In the case
of $^4$He results for two different radii (from \cite{sick82,sick08})
are given. 
\begin{table}[!h]
\begin{center} 
    \begin{tabular}{|lrr|}
      \hline
Contribution  &  Value (meV) &  Uncertainty (meV) \\
      \hline
Uehling        &     1666.305~    &  \\
K\"allen-Sabry     &       11.573~  &           \\
VP iterations \cite{karshenboim}  &  1.709~  &    \\ 
sixth order \cite{karshenboim}  & 0.074~    & 0.003~~~  \\
Total "LBL"  \cite{karshenboim2} &  -0.0136 &  0.0006  \\ 
mixed mu-e VP    &       0.0021   &  \\
hadronic VP       &      0.228~~    &   0.012~~~     \\       
\hline
recoil  \cite{eides} (eq136)     & -0.4330  &  \\
recoil, higher order \cite{eides} & -0.0474 & \\
recoil, finite size \cite{friar79} & 0.2662  &  0.001~ \\
recoil correction  to VP \cite{Karshenboim2012}  &  -0.0090 &  \\
additional recoil  \cite{barker} & 0.2952 &   \\
\hline
muon Lamb shift   &    &  \\
second order      &     -11.1057 &   \\
higher orders      &     -0.0347 &  \\
\hline
nuclear size  ($R_{He}$=1.676\,fm) \cite{sick82}  & &  0.008\,fm \\
main correction $B\cdot\langle r^2 \rangle$  & -298.706~~    &  2.8~~~   \\
Zemach moment \cite{friar79} &  6.591~~  &  0.188~~ \\
remaining order $(\alpha Z)^6$ \cite{friar79} &  0.055~~  &  \\
correction to the 2p$_{1/2}$ level   &   0.0148~  &  \\
\hline
nuclear size  ($R_{He}$=1.681\,fm) \cite{sick08}  & &  0.004\,fm \\
main correction $B\cdot\langle r^2 \rangle$  & -300.491~~    &  1.4~~~   \\
Zemach moment \cite{friar79} &  6.650~~  &  0.190~~ \\
remaining order $(\alpha Z)^6$ \cite{friar79} &  0.056~~  &  \\
correction to the 2p$_{1/2}$ level   &   0.0149~  &  \\
polarization \cite{RMP,friar77}     &    3.1~~ &  ~0.6~~          \\
\hline
\end{tabular}        
 \caption{ Contributions to the muonic helium Lamb shift.  Finite size
   contributions are given for two values of the 
 nuclear radius of $^4$He. }
  \label{tab:Total-He4}
\end{center}   
\vspace{-0.15cm}    
\end{table}        

As shown in  Table\,\ref{tab:coeff} in Appendix~B, the finite
size corrections for $^4$He can be parametrized as 
 \mbox{$-106.340 \langle r^2 \rangle \,+\,0.400 \langle r^3 \rangle_{(2)}$.}  
where energies are in meV and radii in fm.  This parametrization
includes the correction to the 2p$_{1/2}$ level, even though the
numerical value is given separately in the table.     
The second term can be approximated by $1.40(4) (\langle r^2 \rangle)^{3/2}$.
The total transition energy would then be   \\  
\mbox{1668.8641$\pm$0.005+3.1$\pm$0.6-300.477$\pm$1.4
  +6.650$\pm$0.19\,=\,1378.137$\pm$1.5\,meV.}   The nuclear \\
polarization correction has recently been recalculated \cite{ji} to be
 2.47$\pm$0.015\,meV.  This would change the total value.

 In previous work \cite{RMP,Borie78}, recoil corrections to the Lamb
 shift in  $^4$He denoted by "two photon"  and "Breit" were given as 
$-0.44$\,meV and +0.28\,meV, respectively.  These clearly correspond to
corrections given here as (recoil  \cite{eides} (eq136)) of $-0.433$\,meV  
and (recoil, finite size \cite{friar79}) of 
\mbox{0.1221\,fm$^{-1}\langle r \rangle_{(2)} $\,=\,0.266\,meV},
respectively   

For $^3$He   
one finds the  transition energies in meV in Table \ref{tab:Total-He3}.  
Also here the main  vacuum polarization contributions are given for a   
point nucleus, using the Dirac equation with reduced mass. 
\begin{table}[!h]
\begin{center} 
    \begin{tabular}{|lrr|}
      \hline
Contribution  &  Value (meV) &  Uncertainty (meV) \\
      \hline
Uehling        &     1642.412~    &  \\
K\"allen-Sabry     &       11.411~  &           \\
VP iterations \cite{karshenboim2}     &  1.674~  &    \\
sixth order \cite{karshenboim2}  & 0.073~    & 0.003~~~  \\
Total "LBL"  \cite{karshenboim2} &  -0.0134 &  0.0006  \\ 
mixed mu-e VP    &       0.0020   &  \\
hadronic VP       &      0.221~    &   0.011~~~     \\       
\hline
recoil  \cite{eides} (eq136)     & -0.55811  &  \\
recoil, higher order \cite{eides} & -0.0810~ & \\
recoil, finite size \cite{friar79} & 0.4040~  &  0.0010~~ \\
recoil correction  to VP \cite{Karshenboim2012}  &  -0.0140 &  \\
additional recoil  \cite{barker} & 0.12654 &   \\
\hline
muon Lamb shift   &    &  \\
second order      &     -10.8274 &   \\
higher orders      &     -0.0337 &  \\
\hline
nuclear size  ($R_{He}$=1.966\,fm)  &  &  0.010~~~ \\
main correction $B\cdot\langle r^2 \rangle$  & -400.075~    & ~3.67~~~    \\
Zemach moment \cite{friar79} &  10.258~  &  0.305~~~ \\
remaining order $(\alpha Z)^6$ \cite{friar79} &  0.121~  &  \\

correction to the 2p$_{1/2}$ level   &   0.0158~  &  \\
polarization \cite{RMP}     &    4.9~~~~ &  ~~1.0~~~          \\
\hline
\end{tabular}        
 \caption{ Contributions to the muonic helium Lamb shift. The 
 nuclear  radius of $^3$He is taken to be 1.966(10)\,fm. }
  \label{tab:Total-He3}
\end{center}
\vspace{-0.15cm}       
\end{table}    
 The value of the term for "VP iterations" given here differs from the
 value of 1.4\,meV calculated by Rinker \cite{RMP}. From
 Table\,\ref{tab:coeff} in Appendix~B, the finite size corrections for
 $^3$He can be parametrized as 
\mbox{$-103.507(5) \langle r^2 \rangle \,+\,0.3860 \langle r^3 \rangle_{(2)}$.}  
where energies are in meV and radii in fm. 
  The second term can be
approximated by $1.35(4) (\langle r^2 \rangle)^{3/2}$.
This gives a total value of 
(1644.9169$\pm$0.6-103.507(5)$\langle r^2 \rangle\,+\,0.3860\langle r^3 \rangle_{(2)}$ +4.9$\pm$1.0)\,meV.
  Here the estimated uncertainty does not include the uncertainty in the radius,
but only the uncertainty in the coefficient of the Zemach moment and the
nuclear polarization. 
 This agrees quite well with a previous calculation
\cite{BorieHe3}, which was considerably less precise, when the change in
the contribution due the more recent measured radius is taken into account.

\subsubsection*{Fine structure of the 2p state }
\vspace{-0.2cm}
The fine structure of the 2p states can be calculated  by using
the relativistic Dirac energies, computing the vacuum polarization
contributions with Dirac wave functions, and including the effect of the
anomalous magnetic moment in the muon Lamb shift.  
An additional recoil correction (Eq.\ref{eq:recoil1}) also has to be
included.  
The results are summarized in Table\,\ref{tab:FS}.    
\begin{table}[!h]
\begin{center} 
    \begin{tabular}{|lrrrr|}
      \hline
          &  Hydrogen  &   deuterium  &  $^3$He&  $^4$He \\
Dirac          &      8.41564   &  8.86430  & 144.4157 &  145.7183        \\
Uehling(VP)        &     0.00501    &  0.00575 & 0.2696  &  0.2753 \\
K\"allen-Sabry      &     0.00004  &   0.00005 & 0.0021  &  0.0021  \\
\hline
anomalous moment $a_{\mu}  $ &     &  &    \\
second order      &      0.01757   &  0.01949 & 0.3232 & 0.3290 \\
higher orders      &     0.00007   &  0.00007  & 0.0012   &  0.0013 \\
\hline
Recoil  (Eq.(\ref{eq:recoil1}))  &  -0.08621  & -0.02521& -0.1898 & -0.1107  \\
\hline
Finite size   &     -0.00004  &   -0.00027   & -0.0158  &     -0.0119   \\
Total Fine Structure    &      8.35208  &  8.86419  & 144.8062  &  146.2034 \\
\hline
\end{tabular}        
 \caption{ Contributions to the fine structure 
($ E(2p_{3/2}) - E(2p_{1/2})$) of the 2p-state in muonic
   hydrogen, deuterium, $^3$He and $^4$He, in meV.  }
  \label{tab:FS}
\end{center}   
\end{table} 
 One should also include the 
$B^2/2M_N$-type correction to the fine structure.  (see \cite{eides},
Eq(38)).  This is tiny 
($5.7 \cdot 10^{-6}$\,meV for hydrogen) and is not included in the table. 
As mentioned before,  Friar
\cite{friar79} has given  expressions for the energy shifts of the
2p-states due to finite nuclear size.  A  correction proportional to 
$\langle r^2 \rangle$  affects only the $2p_{1/2}$ state, and thus
contributes to the fine structure.  The contribution  to the fine
structure of the 2p-state of hydrogen ($4 \cdot 10^{-5}$\,meV) is negligible.  
However, this contribution is not negligible for the helium isotopes.
For hydrogen,   
this result for the fine structure was subsequently reproduced by Martynenko
\cite{martynenko08}. His results for the hyperfine structure of the 2p
levels in hydrogen also agree with the results given below (and
previously \cite{Borie05}).  For deuterium, the fine structure 
agrees with that given by Krutov and Martynenko \cite{martynenko11}.

\subsubsection*{Hyperfine Interactions }
\vspace{-0.2cm}
The Breit equation  \cite{eides,hyperfine,barker} contributions to the
fine- and hyperfine interactions for general potentials and 
arbitrary spins were given by Metzner and Pilkuhn \cite{rmetzner}.  Here 
a version applicable to the case of muonic atoms ($Z_1=-1$, $s_1=1/2$,
$m_1=m_{\mu}$,  $\kappa_1=a_{\mu}$, $Z_2=Z$) is given. 
For most of the following, the potential is approximated by
\mbox{$V_{Coul}=-\alpha Z/r$.} 
\begin{equation}
 V_{L,s_1} =  \frac{1}{2 m_{\mu}}\, \frac{1}{r}\,\frac{dV}{dr} 
\Bigl[ \frac{1+a_{\mu}}{s_1 m_r} - \frac{1}{m_{\mu}} \Bigr]
 \vec{L} \cdot \vec{s}_{1}
\label{eq:fs1}
\end{equation}%
This can be rearranged to give the well-known form for spin 1/2
nuclei with an anomalous magnetic moment, namely  
\begin{equation*}
- \frac{1}{r}\,\frac{dV}{dr} \cdot \frac{1+a_{\mu}+(a_{\mu}+1/2)m_N/m_{\mu}}
{m_N m_{\mu}} \vec{L} \cdot \vec{\sigma}_{\mu}
\end{equation*}%
Note that 
\begin{equation*}
 \frac{1}{m_N m_{\mu}}+\frac{1}{2 m_{\mu}^2}\,=\,
  \frac{1}{2 m_r^2}-\frac{1}{2 m_N^2}
\end{equation*}
so that the terms not involving  $a_{\mu}$ in the spin-orbit
contribution are really the Dirac fine
structure plus the Barker-Glover correction (Eq. \ref{eq:recoil1}).

Also
\begin{equation*}
 V_{L,s_2} =  \frac{1}{2 m_{2}}\, \frac{1}{r}\,\frac{dV}{dr} 
\Bigl[ \frac{1+\kappa_{2}}{s_2 m_r} - \frac{1}{m_{2}} \Bigr]
 \vec{L} \cdot \vec{s}_{2}
\end{equation*}%
Usually one writes 
\begin{equation*}
  \frac{Z(1+\kappa_{2})}{ m_2} = \frac{\mu_2}{m_{p}} 
\end{equation*} 
\noindent   where $\mu_2$ is the magnetic moment of the nucleus in units
of nuclear magnetons \mbox{($\mu_N=e/2 m_p$).}  A value of 
$\mu_d\,=\,0.85744\,\mu_N\,=\,0.307012\,\mu_p$ corresponds to
$\kappa_d$\,=\,0.714.   In the case of $^3$He, the magnetic moment is 
$\mu_{He3}\,=\,-2.1275\,\mu_N$, resulting in $\kappa_{He3}\,=\,-4.185$ 

\noindent The spin-spin interaction is given by
\begin{equation*}
 V_{s_1,s_2} = \frac{2 (1+a_{\mu})\mu_2}{2 s_2 m_{\mu} m_{2}}\, 
 \Bigl[ \frac{1}{r}\,\frac{dV}{dr} 
(3 \vec{s}_1 \cdot \hat{r} \vec{s}_2 \cdot \hat{r} - \vec{s}_1
 \cdot \vec{s}_{2}) - \frac{2}{3} \nabla^2 V \vec{s}_1 \cdot \vec{s}_{2}
 \Bigr]  
\end{equation*} 

For deuterium the quadrupole moment also contributes, with 
\begin{equation*}
 V_{Q} = - \frac{Q}{2} \,  \frac{1}{r}\,\frac{dV}{dr} 
\Bigl[3 \vec{s}_2 \cdot \hat{r} \vec{s}_2 \cdot \hat{r} - \vec{s}_2
 \cdot \vec{s}_{2} \Bigr] 
\end{equation*}%
\noindent  with Q in units of $1/m_2^2$.  
 The quadrupole moment of the deuteron is taken to be
Q\,=\,0.2860(15)\,fm$^2$~\cite{friar02,reid-v,bishop}.  In other units,
one finds  Q\,=\,25.84/$m_d^2$\,=\,7.345$\times10^{-6}$\,MeV$^{-2}$. 

The Uehling potential has to be included in the potential $V(r)$.  
For states with $\ell\,>\,0$ in light atoms, and
neglecting the effect of finite nuclear size, we may take 
\begin{equation}
\frac{1}{r} \frac{d V}{dr}~=~\frac{ \alpha Z}{r^3} \cdot \left[ 1 +
 \frac{2 \alpha}{3 \pi} \int_1^{\infty} \frac{(z^2-1)^{1/2}}{z^2}\cdot
\left(1+\frac{1}{2 z^2}\right) \cdot (1 + 2 m_e r z) \cdot e^{-2 m_e r z} \,dz \right]
\label{eq:vp2p}
\end{equation}
which is obtained from the Uehling potential \cite{RMP,Uehling,serber}
by differentiation.  Then, assuming that it is sufficient to use
nonrelativistic point Coulomb wave functions for the 2p state, one finds 
\begin{equation*}
\Big\langle \frac{1}{r} \frac{d V}{dr} \Big\rangle_{2p} \rightarrow 
\Big\langle \frac{1}{r}\frac{d V}{dr}\Big\rangle_{2p}\cdot (1+\varepsilon_{2p})
\end{equation*}
\noindent  where 
\begin{equation}
\varepsilon_{2p} ~=~ \frac{2 \alpha}{3 \pi} 
 \int_1^{\infty} \frac{(z^2-1)^{1/2}}{z^2}\cdot \left(1+\frac{1}{2 z^2}\right)
  \cdot \left(\frac{1}{(1+az)^2} + \frac{2 az}{(1+az)^3}\right) \,dz
\label{eq:eps2p}
\end{equation}
\noindent  with $a \,=\,2 m_e /(\alpha Z m_r)$.  
For hydrogen,  $\varepsilon_{2p}$\,=\,0.000365,  for deuterium 
$\varepsilon_{2p}$\,=\,0.000391 and for 
$^3$He $\varepsilon_{2p}$\,=\,0.000894.  

The correction due to vacuum polarization (Eq.\,(\ref{eq:eps2p})) should be
applied to the HFS shifts of the 2p-states, (and also to all  spin-orbit
terms in the "muon Lamb shift").    

Note that for the 2p state 
\begin{equation*}
  \Big\langle \frac{1}{r}\,\frac{dV}{dr} \Big\rangle \,=\, 2\alpha Z 
  \frac{(\alpha Z m_r/n)^3}{\ell(\ell+1)(2\ell+1)} \,=\, 
 \frac{\alpha Z (\alpha Z m_r)^3}{24} 
\end{equation*}%

 The hyperfine structure can also be calculated relativistically,
 using the formalism given in \cite{RMP}.  The finite extent of the
 magnetization density and the effect of vacuum polarization should be
 taken into account also in this approach.  


\subsubsection*{Hyperfine structure of the  2s-state:}
\vspace{-0.2cm}
The expectation value of $V_{s_1 s_2}$ in an n-s state with $j=1/2$ is 
\begin{equation*}
\Delta E_{ns}=\,=\,\frac{2 \mu_2 \alpha (\alpha Z)^3 m_r^3}
{3 n^3 m_{\mu} m_2 s_2} \cdot (1 + a_{\mu})  [F(F+1)-s_2(s_2+1)-3/4]
\end{equation*}
When $s_2=1/2$, and $\mu_2/m_p=(1+\kappa_2)/m_2$, this reproduces the
well-known result for muonic hydrogen (see, for example \cite{eides},
Eq. (271,277): 
\begin{equation*}
\Delta E_{ns}\,=\,\frac{2}{3 m_{\mu} m_2} \cdot (1 + \kappa_2) \cdot (1 + a_{\mu})
 \langle \nabla^2 V \rangle [F(F+1)- 3/2] 
 \,=\, \frac{\beta}{2} \cdot (1 + a_{\mu})  [F(F+1)- 3/2] 
\end{equation*}
\noindent with  
\begin{equation}
 \beta \,=\,\frac{8(\alpha Z)^4 m_r^3}{3 n^3 m_{\mu} m_2} \cdot (1 + \kappa_2) 
 \,=\, (8/n^3) \times 22.8054\,meV 
\label{eq:beta}
\end{equation}
\noindent
The numerical value was calculated for hydrogen.  Note that $\beta \cdot
(1 + a_{\mu})$\,=\,22.8320\,meV, which is the well-known value. 
Since $^3$He also has spin 1/2, the same
formula is valid, and one obtains $\beta$\,=\,-171.3964\,meV.
($\beta \cdot (1 + a_{\mu})$\,=\,-171.5963\,meV)

For deuterium, with $s_2=1$, the corresponding hyperfine splitting is 
\begin{equation*}  \begin{split}
\Delta  E_{ns} &\,=\,  \frac{2 (\alpha Z)^4 m_r^3}{3 n^3 m_{\mu} m_D}
 \cdot (1 + \kappa_D) \cdot (1 + a_{\mu}) \cdot [F(F+1) - 11/4] \\
  & \,=\, \frac{\beta_D}{2} \cdot (1 + a_{\mu}) \cdot [F(F+1) - 11/4] 
   \,=\, (8/n^3) \cdot 2.04766\,meV \cdot [ \delta_{F,3/2} - 2  \delta_{F,1/2}]
\end{split}
\end{equation*}
\noindent for a total splitting of 6.14298\,meV in muonic deuterium.  
This is in reasonably good agreement with the result given by 
Carboni \cite{carboni}.

The QED corrections to the energy shift of the 2s-state were discussed
in \cite{eides,hyperfine}.  In   muonic hydrogen they are  given by: 
\begin{equation}
\Delta E_{2s}~=~\beta \cdot (1 + a_{\mu}) \cdot (1+\varepsilon_{VP} +
 \varepsilon_{vertex} + \varepsilon_{Breit} + \varepsilon_{Zem}) 
 \cdot [\delta_{F1}-3\delta_{F0}]/4  
\label{eq:hf2s}
\end{equation}
\noindent The corrections due to QED effects, nuclear size and recoil
 are analogous for muonic deuterium and $^3$He.    Other corrections due to
 recoil, nuclear polarization, the weak interaction and so on can be
 included by  adding corresponding  $\varepsilon$'s.     
 Here  (\cite{eides}, Eq.\,(277))
\begin{equation*}
\varepsilon_{Breit} \,=\, \frac{17 (\alpha Z)^2}{8} 
  \,=\, 1.13 \cdot 10^{-4} \cdot Z^2
\end{equation*}
\noindent  is a relativistic correction that gives the difference
between the value obtained in relativistic perturbation theory using  Dirac
wave functions, and the nonrelativistic value  (to leading order in
$(\alpha Z)^2$). 
The vertex correction  (\cite{eides,Brodsky}) is given by 
\begin{equation*}
\varepsilon_{vertex} \,=\,  \alpha (Z \alpha ) 
\left(\ln(2)-\frac{5}{2}\right) \,=\, -0.9622 \cdot 10^{-4} \cdot Z
\end{equation*}
This includes a correction of 
$3 \alpha (Z \alpha )/4  \,=\, 0.3994 \cdot 10^{-4} \cdot Z $
due to muon loop  vacuum polarization. Corrections due to hadronic
vacuum polarization can be expected to be comparable to this value. 
 An estimate following the prescription given in \cite{hadron} gives 
a correction of approximately $0.2666 \cdot 10^{-4} \cdot Z $, or 
approximately 0.0006(1)\,meV in hydrogen.
Some higher order corrections to $\varepsilon_{vertex}$ of order $\alpha
(Z \alpha )^2 \ln^2(Z \alpha)$  are possibly numerically important and
were given in Ref.\,\cite{Brodsky} as
\begin{equation*}
  \frac{\alpha (Z \alpha)^2} {\pi}
\left(-\frac{2}{3}\ln^2((Z \alpha)^{-2})+c_{22}(2)\ln((Z \alpha)^{-2}) \right)
 \,=\, -0.0073 \cdot 10^{-4}           \qquad  (Z=1)
\end{equation*}
This correction adds an additional -0.00017\,meV to the hyperfine
splitting in the  case of hydrogen (and -0.00004\,meV in muonic deuterium).
If Z=2, the numerical value of the extra correction is $-0.211\times 10^{-4}$
 and the contribution to the hyperfine splitting in muonic $^3$He is  
 -0.0036\,meV. 

The main vacuum polarization correction has two contributions.  One of 
these is a result of a modification of the magnetic interaction between the
muon and the nucleus and is given by (see \cite{BorieHe3})
\begin{equation} \begin{split}
\varepsilon_{VP1} \,=\,& \frac{4 \alpha}{3 \pi^2}  
 \int_0^{\infty} r^2 \,dr \left(\frac{R_{ns}(r)}{R_{ns}(0)}\right)^2
 \int_0^{\infty} q^4 j_0(qr) G_M(q) \,dq   \\
 & \int_1^{\infty} \frac{(z^2-1)^{1/2}}{z^2}\cdot
\left(1+\frac{1}{2 z^2}\right) \cdot \frac{dz}{4 m_e^2 [z^2 + (q/2 m_e)^2]} 
\end{split}
\label{eq:epsvp}
\end{equation}
One can do two of the integrals analytically and obtains for the
2s-state (with
\mbox{$a=2m_e/(\alpha Z m_r)$} and  \mbox{$\sinh(\phi) = q/(2 m_e) = K/a$})  
\begin{equation}
\varepsilon_{VP1} \,=\, \frac{4 \alpha}{3 \pi^2}  
 \int_0^{\infty} \frac{K^2}{(1+K^2)^2} F(\phi) G_M(\alpha Z m_r K) \,dK 
\left[2 -\frac{7}{(1+K^2)}+\frac{6}{(1+K^2)^2}\right]
\label{eq:epsvp1}
\end{equation}
where $F(\phi)$ is known from the Fourier transform of the Uehling
potential and is given by Eq(\ref{eq:f-phi}).  

The other contribution, as discussed by \cite{Brodsky,sternheim} arises
from the 
fact that the lower energy hyperfine state, being more tightly bound,
has a higher probability of being in a region where vacuum polarization
is large.  This results in an additional energy shift of  
\begin{equation*}
 2 \int V_{Uehl}(r) \psi_{2s}(r) \delta_M \psi_{2s}(r) d^3r
\end{equation*}
Following Ref.\,\cite{Brodsky} with $y=(\alpha Z m_r/2) \cdot r$, one has 
\begin{equation}
   \delta_M \psi_{2s}(r) \,=\, 2\alpha Z m^2 \Delta \nu_F \psi_{2s}(0) 
\left(\frac{2}{\alpha Z m_r}\right)^3  e^{-y} 
\left[(1-y)(\ln(2y)+\gamma)+\frac{13y-3-2y^2}{4}-\frac{1}{4y}\right]
\label{eq:delta-psi}
\end{equation}
\noindent ($\gamma$ is Euler's constant), and
\begin{equation*}
  \psi_{2s}(r)\, =\, \psi_{2s}(0) (1-y)  e^{-y} 
\end{equation*}
 One finds after a lengthy integration
\begin{multline}
\varepsilon_{VP2} \,=\, \frac{16 \alpha}{3 \pi^2} 
 \int_0^{\infty} \frac{dK}{1+K^2}  G_E(\alpha Z m_r K)  F(\phi) \\
\biggl\{
 \frac{1}{2}-\frac{17}{(1+K^2)^2}+\frac{41}{(1+K^2)^3}-\frac{24}{(1+K^2)^4} \\
~+\frac{\ln(1+K^2)}{1+K^2} \left[2-\frac{7}{(1+K^2)}+\frac{6}{(1+K^2)^2}\right]
 \\  + \frac{\tan ^{-1}(K)}{K}
 \biggl[1-\frac{19}{2(1+K^2)}+\frac{20}{(1+K^2)^2}-\frac{12}{(1+K^2)^3}\biggr] 
 \biggr\}
\label{eq:epsvp2}
\end{multline}
Sternheim\cite{sternheim} denotes the two contributions by $\delta_M$
and  $\delta_E$, respectively.  Note that the latter contribution is
affected by the  nuclear charge distribution.  
An alternative exression, obtained by assuming a point nucleus, using
Eq.(131)  from \cite{RMP} for the Uehling potential, and doing the
integrations in  a different order, is  
\begin{equation} \begin{split}
\varepsilon_{VP2} \,=\,& \frac{16 \alpha}{3 \pi} 
 \int_1^{\infty} \frac{(z^2-1)^{1/2}}{z^2}\cdot
  \left(1+\frac{1}{2 z^2}\right) \cdot \frac{1}{(1+az)^2}   \\
& \cdot \biggl[\frac{az}{2}-\frac{1}{1+az}+\frac{23}{8(1+az)^2}-
 \frac{3}{2(1+az)^3}    \\ 
&  +\ln(1+az)\cdot\left(1-\frac{2}{1+az}+\frac{3}{2(1+az)^2}\right)\biggr] 
\,dz 
\end{split}
\end{equation}
\noindent  with $a \,=\,2 m_e /(\alpha Z m_{red})$ as in the case of
 $\varepsilon_{2p}$.  Both methods give the same result when the effect
 of nuclear size is neglected. 
In the case of ordinary hydrogen, each of these contributes 
$3 \alpha^2/8 = 1.997 \cdot 10^{-5}$.  The accuracy of the numerical
integration was checked by reproducing these results. 
One can thus expect that muonic vacuum polarization will contribute 
$3 \alpha^2/4 \simeq 4 \cdot 10^{-5}$, as in the case of normal
hydrogen.  This correction was included in  $\varepsilon_{vertex}$. 
Martynenko \cite{martynenko} includes only one of these contributions.  
The energy shift is approximately 0.00091\,meV in hydrogen.
Contributions due to the weak interaction were estimated by Eides
\cite{eides-weak} and give  $5.8\times10^{-8} (m_{\mu}/m_e) E_F$ which works
out to 0.00027\,meV for the weak interaction contribution in muonic hydrogen. 
This corresponds to $\varepsilon_{weak}\,=\,1.2 \cdot 10^{-5}$ for hydrogen.  
Later this work has been extended (\cite{eides-weak2}) with the result that 
$\varepsilon_{weak}\,=\,1.5 \cdot 10^{-5}$ for $^3$He, and for deuterium 
$\varepsilon_{weak}\,\approx \,0$.  


Finite nuclear size has an effect on the value of $\varepsilon_{VP}$.   
For muonic hydrogen, one obtains 
 $\varepsilon_{VP1}$=0.00211 and  $\varepsilon_{VP2}$=0.00326 for a
 point nucleus and  $\varepsilon_{VP1}$=0.00206 and  
 $\varepsilon_{VP2}$=0.00321 with a proton radius of 0.875\,fm.  
 Including this effect reduces the total contribution due to vacuum
polarization by 0.00023\,meV.
 
For muonic $^3$He, one obtains 
 $\varepsilon_{VP1}$=0.00295 and  $\varepsilon_{VP2}$=0.00486 including
 the effect of finite nuclear size.  For a point nucleus the values would
 be  $\varepsilon_{VP1}$=0.00315 and  $\varepsilon_{VP2}$=0.00506. 
 
For the case of muonic deuterium, 
 $\varepsilon_{VP1}$=0.00218 and  $\varepsilon_{VP2}$=0.00337 for a 
point nucleus.  Including the effect of finite nuclear size gives 
$\varepsilon_{VP1}$=0.00207 and  $\varepsilon_{VP2}$=0.00326. This 
has the effect of reducing the total value of the hyperfine splitting by
0.00134\,meV. 

The contribution to the hyperfine splitting of the 2s-state of muonic
hydrogen (with  \\
 finite extension of the magnetization (and charge) density) is \\ 
 0.04703\,meV+0.07328\,meV=0.12031\,meV.   


The contribution to the hyperfine structure from the two loop diagrams     
\cite{kaellen} can be calculated by replacing 
$U_2(\alpha Z m_r K) = (\alpha / 3\pi)  F(\phi) $ by 
$U_4(\alpha Z m_r K)$ (as given in \cite{RMP,Borie75}) in equations 
\ref{eq:epsvp1} and \ref{eq:epsvp2}.  The resulting contributions in
hydrogen are  
$1.636 \cdot 10^{-5}$ and $2.460 \cdot 10^{-5}$, respectively, giving a
total shift of 0.00093\,meV.  Martynenko \cite{martynenko} neglected the
contribution corresponding to  Eq.\,\ref{eq:epsvp2}. 
The contributions in muonic deuterium are $1.688 \cdot 10^{-5}$ and
$2.545 \cdot 10^{-5}$, respectively.  For muonic  $^3$He they are 
$2.511 \cdot 10^{-5}$ and $3.928 \cdot 10^{-5}$, respectively.  

The correction due to $\varepsilon_{VP2}$ can be regarded as a result of
distortion of the wave function of the 2s-state, and has recently been
calculated by Ivanov et al. \cite{ivanov}.  Numerically their results
agree with those given here to within a few tenths of one per cent, for both 
the second order and fourth order contributions in hydrogen and in deuterium. 
A more complete calculation of the vacuum polarization corrections to
the 2s-state in hydrogen by the same authors \cite{karshenboim-hfs}
gives the same  result for all these corrections to three significant
figures, when nuclear size is neglected.  This paper includes the
contribution of three further graphs, which give an additional
contribution to $\varepsilon_{VPtot} $ of $1.64 \cdot 10^{-5}$, and a
total two-loop contribution of 0.00130\,meV to the hyperfine splitting of the
2s-state of hydrogen. 

The main correction due to finite extension of the magnetization density 
is known as the Bohr-Weisskopf effect (\cite{zemach}) and is equal to 
\begin{equation*}
\varepsilon_{Zem} \,=\, -2  \alpha Z m_r \langle r \rangle_{(2)}
\end{equation*}
where  $\langle r \rangle_{(2)}$ is given in \cite{hyperfine,friar79,friar04}.

Ref.\,\cite{Pachuki1}   claims that this correction
does not treat off-shell effects and/or recoil properly.   For hydrogen, 
Carlson et al.~\cite{carlson08} give a recoil correction corresponding to
$\varepsilon_{rec} \,=\,0.00093$, resulting in a correction of 0.02123\,meV.   

For hydrogen the value of $2 \alpha Z m_r $ is 0.007024$fm^{-1}$. 
Using the value \\
\mbox{$\langle r \rangle_{(2)}\,=\,1.086 \pm 0.012$\,fm} from \cite{friar04},
gives a  Zemach correction of  
   $\varepsilon_{Zem}\,=\,-0.00762$, and a contribution of 
 \mbox{-0.1742(19)\,meV} to the hyperfine splitting of the 2s state in hydrogen.
 Distler et al. \cite{newzemach} give 
$\langle r \rangle_{(2)}\,=\,1.045 \pm 0.004$\,fm  for the 'magnetic'
Zemach radius; the resulting value 
 of  $\varepsilon_{Zem}$ is -0.00734, and an energy shift of
 \mbox{-0.1676(6)\,meV}.  

The corrections due to finite size and recoil have been given in
\cite{Pachuki1} as -0.145\,meV, while a value of -0.152\,meV is given in
\cite{martynenko}.  Combining the recoil correction from Ref.~\cite{carlson08}
with the Zemach correction results in values compatible with either of
these, depending on the Zemach radius (for example,  -0.1464(6)\,meV
when the Zemach radius of Distler et al. \cite{newzemach} is used).

A correction for possible nuclear polarization effects has  been
calculated for hydrogen by Cherednikova et al. \cite{hfs-pol1} with the 
result $\varepsilon_{pol}\,=\,0.00046(8)$, for an additional contribution of
 $0.0105\pm 0.0018$\,meV.  Carlson et al.~\cite{carlson08} give a value
 of  $\varepsilon_{pol}\,=\,0.000351(114)$, or $0.0080\pm 0.0026$\,meV for 
this correction.   To be consistent with the recoil correction given
above, the correction of Carlson et al. should be used (see also
ref.\,\cite{carlson11a}).   

A comparable calculation for muonic  $^3$He has not been found.   
A very recent preprint by Faustov et al.~\cite{Faustovetal14} gives a
result for the polarizability correction in  muonic deuterium of
0.2226\,meV.  It is not clear whether the "elastic" contribution of the
two-photon exchange diagrams is taken into account as described in
Refs.~\cite{carlson08,carlson11a}.  



In hydrogen, the total corrections (other than for the Zemach radius)
amount to   \\
$0.1533\pm0.0027$\,meV  (or $0.1510\pm0.0027$\,meV if the finite extent
of the magnetization density in the VP correction is included).  
Adding $E_F$ gives 22.9853$\pm0.0027$\,meV. 
The correction due  to the Zemach radius is 
\mbox{-0.16037\,meV\,fm$^{-1}\langle r \rangle_{(2)}$.}   
If the value of the Zemach radius from  Distler et al. \cite{newzemach}
is used, the total hyperfine splitting in the 2s-state of muonic
hydrogen is 22.8177(30)\,meV.    
This is in substantial agreement with the recent results given in
Ref.\,\cite{Indelicato}, 

Alternatively, one could replace the contribution from the Zemach
correction by a polynomial expansion in the magnetic radius, as
suggested by Carroll et al.\cite{carroll}.  Since the leading term in a
perturbative calculation (relativistic or nonrelativistic) is linear, 
this should take the form ($a r + b r^2$).  Since these authors take
most of their radiative corrections from other work \cite{martynenko},
which neglects some of the corrections included here, a more detailed
comparison is not appropriate. 

 For muonic deuterium,  The coefficient of 
$\langle r \rangle_{(2)}$ is -0.007398\,fm$^{-1}$, giving, with  \\
\mbox{$\langle r \rangle_{(2)}\,=\,2.593 \pm 0.016$\,fm} from \cite{friar04},
$\varepsilon_{Zem}\,=\,-0.01918 \pm 0.00012$.  Nuclear polarization
and recoil corrections are important, but have not been included. 
The total hyperfine splitting of the 2s-state, including all corrections
other than the possible polarization correction, is  
\begin{equation*}
\Delta E_{2s}~=~ \frac{3}{2} \beta_D \cdot (1 + a_{\mu}) \cdot
 (1+\varepsilon_{VP} + \varepsilon_{vertex} + \varepsilon_{Breit} +
 \varepsilon_{Zem}) \,=\,6.0584(7)\,meV     
\end{equation*}
If  the effect of finite nuclear size on the vacuum polarization
corrections is not included, the result would be 6.0597\,meV. 

For muonic $^3$He,  The coefficient of 
$\langle r \rangle_{(2)}$ is -0.01506\,fm$^{-1}$, giving, with 
\mbox{$\langle r \rangle_{(2)}\,=\,2.562$ \,fm}  (a Gaussian charge
distribution was assumed), $\varepsilon_{Zem}\,=\,-0.0386$.
The total hyperfine splitting of the 2s-state, including all known 
corrections, is  -166.3745\,meV.

\subsubsection*{Hyperfine structure of the 2p state}
\vspace{-0.2cm}
The hyperfine structure of muonic hydrogen is calculated in the same way
as was done in 
earlier work \cite{hyperfine,BorieHe3}, but with improved accuracy.  
Most of the formalism and results are similar to those given by
\cite{Pachuki1} and \cite{brodsky-p}. 
\subsubsection*{Hydrogen}
\vspace{-0.2cm}
The hyperfine structure of the 2p-state in hydrogen is given by
\cite{hyperfine,brodsky-p} ($F$ is the total angular momentum of the state)
\begin{equation} \begin{split}
\frac{1}{4m_{\mu} m_N} & \Big\langle \frac{1}{r}\,\frac{dV}{dr}\Big\rangle_{2p}
  \cdot(1+\kappa) \biggl[2(1+x) \delta_{j j'}  [F(F+1)-j(j+1) + 3/4] \\
 & + 6 \hat{j} \hat{j}' (C_{F1}(1+a_{\mu})-2(1+x)) \left\{ \begin{array}{ccc} 
 \ell & F & 1 \\
 \frac{1}{2} & \frac{1}{2} & j 
 \end{array} \right\}  \left\{  \begin{array}{ccc} 
 \ell & F & 1 \\
 \frac{1}{2} & \frac{1}{2} & j' 
 \end{array} \right\}     \biggr] 
\end{split}
\end{equation}
\noindent where $\hat{j} = \sqrt{2 j + 1}$, the 6-j symbols are defined
 in \cite{edmonds},  and \\
 $C_{F1}=\delta_{F1}-2\delta_{F0}-(1/5)\delta_{F2}$ 

Also 
\begin{equation*}
 x\,=\, \frac{m_{\mu} (1 + 2 \kappa)}{2 m_N (1 + \kappa)}
\end{equation*}
\noindent  represents a recoil correction due to Thomas precession
\cite{hyperfine,barker,brodsky-p}.  For muonic hydrogen, $x=0.0924$. 

As has been known for a long time \cite{Pachuki1,hyperfine,BorieHe3,brodsky-p},
the states with total angular momentum $F=1$ are a superposition of the
states with $j=1/2$ and $j=3/2$.  
Let the fine structure splitting be denoted by 
$\delta \,=\, E_{2p3/2} - E_{2p1/2} $, and let 
 $\beta ' \,=\, \beta \cdot (1+ \varepsilon_{2p})$, to
take the correction due to vacuum polarization into account.
($\beta$ was defined above, in Eq.\,\ref{eq:beta}).
For hydrogen, $\varepsilon_{2p}$\,=\,0.000365 and  $\beta'$\,=\,22.8138\,meV 

The energy shifts of the 2p-states with total angular momentum F
(notation $^{2F+1}L_j$) are then given in Table \ref{tab:HFS-2p} 
\begin{table}[!h]
\begin{center} 
  \begin{tabular}{|lrr|}
 \hline
State  &          Energy         &   Energy in meV  \\
     \hline
$^1 p_{1/2}$    & -$\beta ' (2+x + a_{\mu})/8 $~    &   -5.9704       \\
$^3 p_{1/2}$    & $(\Delta - R)/2$~    &     1.8458        \\
$^3 p_{3/2}$    & $(\Delta + R)/2$~     &    6.3760         \\
$^5 p_{3/2}$    & $\delta + \beta ' (1+5x/4 - a_{\mu}/4)/20$   & 9.6242   \\
 \hline
\end{tabular}        
 \caption{ Hyperfine  structure of the 2p-state in muonic
   hydrogen. Here $\delta$ = 8.352\,meV is the fine structure splitting
   of the 2p state. }
  \label{tab:HFS-2p}
\end{center}   
\end{table}    
\vspace{-0.1cm}
\begin{center}     
\vspace{-0.1cm}
 \begin{minipage}{\linewidth}
\noindent   where
\begin{equation*}
\Delta\,=\, \delta - \beta '(x - a_{\mu}) /16
\end{equation*}
\begin{equation*}
 R^2 \,=\, [\delta  - \beta '(1 +7x/8 + a_{\mu}/8) /6]^2
 + (\beta')^2 (1+2x-a_{\mu})^2/288  
\end{equation*}
\end{minipage}
\end{center}   
\vspace{-0.1cm}
Some minor errors in \cite{hyperfine} have been corrected.  These
numbers differ slightly from those given in ref.\,\cite{eides}. 

\subsubsection*{Helium-3}
\vspace{-0.2cm}
The formulas for muonic $^3$He are identical to those for hydrogen, but
the numerical values are, of course, different.  
\begin{table}[!h]
\begin{center} 
  \begin{tabular}{|lrr|}
 \hline
State  &          Energy         &   Energy in meV  \\
     \hline
$^1 p_{1/2}$    & -$\beta ' (2+x + a_{\mu})/8 $~    &   43.8458       \\
$^3 p_{1/2}$    & $(\Delta - R)/2$~    &    -14.7877        \\
$^3 p_{3/2}$    & $(\Delta + R)/2$~     &    160.0510         \\
$^5 p_{3/2}$    & $\delta + \beta ' (1+5x/4 - a_{\mu}/4)/20$   & 135.7673   \\
 \hline
\end{tabular}        
 \caption{ Hyperfine  structure of the 2p-state in muonic
   $^3$He. Here $\delta$ = 144.809\,meV. }
  \label{tab:heliumHFS-2p}
\end{center}   
\end{table}    
 For muonic  $^3$He, $x=0.0435$, 
$\beta$ \,=\,-171.396\,meV   $\varepsilon_{2p}$=0.000894,  and 
$\beta ' \,=\,\beta \cdot (1+ \varepsilon_{2p})$ \,=\,-171.550\,meV.  
The energy shifts of the 2p-states with total angular momentum F
(notation $^{2F+1}L_j$) are then given in Table\,\ref{tab:heliumHFS-2p}.

\subsubsection*{Deuterium}
\vspace{-0.2cm}
For the 2p state, the matrix elements of the magnetic hyperfine
structure have been given by Brodsky and Parsons \cite{brodsky-p}.    
For hydrogen they are the same as those calculated above.

The matrix elements for the magnetic hyperfine structure are then given
by \\
\begin{center} 
    \begin{tabular}{|ccl|}
\hline
 $j$  &    $j'$      &   Energy           \\
     \hline
1/2   & 1/2   &  $(\beta_D/6) (2+x_d+a_{\mu})[ -\delta_{F,1/2} + 1/2
    \, \delta_{F,3/2}]   $           \\
3/2  &  3/2   & $\delta\,+\,(\beta_D/4)(4+5x_d -a_{\mu}) [-1/6 \,
     \delta_{F,1/2} - 1/15 \,\delta_{F,3/2} + 1/10 \,\delta_{F,5/2} ]$     \\
3/2  & 1/2  &  $(\beta_D/48) (1 + 2x_d -a_{\mu}) [- \sqrt{2} \,\delta_{F,1/2}  
   - \sqrt{5}\, \delta_{F,3/2} ]$             \\
     \hline
\end{tabular}        
\end{center}   
where $\beta_D$ was defined in the previous section.  Numerically it is
equal to 4.0906\,meV. 
 $x_d\,=\,(m^2_{\mu}/ m_{d} m_r)(\kappa_d/(1+\kappa_d))$\,=\,0.0248. 

For the evaluation of the contributions of the quadrupole HFS, let 
\begin{equation*}
 \epsilon_Q = \frac{Q}{2} \, \Big\langle \frac{1}{r}\,\frac{dV}{dr} \Big\rangle
\end{equation*}%

Numerically one finds, for a point Coulomb potential, and the 2p-state, 
\begin{equation*}
\epsilon_Q \,=\, \frac{\alpha Q}{2} \frac{(\alpha Z m_r)^3}{24}\,=\,0.43423\,meV.  
\end{equation*}  

As mentioned before, the Uehling potential has to be included in the
potential $V(r)$.  For states with  $\ell\,>\,0$ in light atoms, this
can be taken into account by multiplying $\beta_D$ and $\epsilon_Q$ by  
(1+ $\varepsilon_{2p}$) where $\varepsilon_{2p}$ is given by
Eq.(\ref{eq:eps2p}). 
With a numerical value of $\varepsilon_{2p}$\,=\,0.000391 for muonic
deuterium, the value of $\epsilon_Q$ is increased to 0.43439\,meV and
the value of $\beta_D$ is increased to $\beta'_D$=4.0922\,meV.    

The quadrupole interaction results in energy shifts of 
\begin{center} 
    \begin{tabular}{|ccl|}
 \hline
 $j$  &    $j'$      &   Energy           \\
     \hline
1/2   & 1/2   &  $0$           \\
3/2  &  3/2   & $ \epsilon_Q\,[\delta_{F,1/2} - 4/5 \,\delta_{F,3/2} + 
   1/5 \, \delta_{F,5/2} ]$~                \\
3/2  & 1/2  &  $ \epsilon_Q\,[ \sqrt{2} \, \delta_{F,1/2} - 
   1/\sqrt{5} \,\delta_{F,3/2} ]$             \\
     \hline
\end{tabular}        
\end{center}

Then for the 2p-level with $j=j'=3/2$ and $F=5/2$, the energy shift is
given by
\mbox{$\delta\,+\,\epsilon_Q/5\,+\,(\beta_D/40)(4+5x_d-a_{\mu})$\,=\,9.3728\,meV.}    
For the 2p-levels with  $F=1/2$ and  $F=3/2$, the corresponding matrices
have to be diagonalized.    The resulting numerical values for the 
eigenvalues are given in Table\,\ref{tab:HFS-D-2p}.
\begin{table}[!h]
\begin{center} 
  \begin{tabular}{|lr|}
 \hline
State  &           Energy in meV  \\
     \hline
$^2 p_{1/2}$        &   -1.4056       \\   
$^2 p_{3/2}$        &   8.6194         \\  
$^4 p_{1/2}$      &     0.6703        \\
$^4 p_{3/2}$     & 8.2560   \\
$^6 p_{3/2}$     & 9.3728   \\    
 \hline
\end{tabular}        
 \caption{ Hyperfine  structure of the 2p-state in muonic
   deuterium. Here \mbox{$\delta$ = 8.86419\,meV.} }
  \label{tab:HFS-D-2p}
\end{center}   
\end{table}    

\vspace{-0.1cm}

Table  \ref{tab:HFS-mu-H} gives the contributions
to the transition energies due to fine and hyperfine structure in muonic
hydrogen  relative to the 2s-2p$_{1/2}$ transition energy given in
Table\,\ref{tab:final}.
\begin{table}[!h]
\begin{center} 
    \begin{tabular}{|lr|}
     \hline
Transition   &   Energy shift in meV  \\
     \hline
$^1 p_{1/2} - ^3 s_{1/2}$        &   -11.6748       \\   
$^3 p_{1/2} - ^1 s_{1/2}$       &    18.9591        \\  
$^3 p_{1/2} - ^3 s_{1/2}$       &    -3.8586        \\   
$^3 p_{3/2} - ^1 s_{1/2}$         &  23.4893         \\   
$^3 p_{3/2} - ^3 s_{1/2}$         &    0.6716         \\  
$^5 p_{3/2} - ^3 s_{1/2}$         &   3.9198         \\   
     \hline
\end{tabular}        
 \caption{ Fine- and hyperfine contributions to the Lamb shift in muonic
   hydrogen.   }
  \label{tab:HFS-mu-H}
\end{center}   
\end{table}    

Table  \ref{tab:HFS-mu-d} gives the contributions
to the transition energies due to fine and hyperfine structure in deuterium.
\begin{table}[!h]
\begin{center} 
    \begin{tabular}{|lr|}
     \hline
Transition   &   Energy shift in meV  \\
     \hline
$^2 p_{1/2} - ^2 s_{1/2}$        &   2.6333       \\
$^2 p_{3/2} - ^2 s_{1/2}$       &    12.6583        \\
$^4 p_{1/2} - ^2 s_{1/2}$       &     4.7092        \\
$^4 p_{3/2} - ^2 s_{1/2}$         &  12.2949         \\
$^2 p_{1/2} - ^4 s_{1/2}$        &   -3.4251       \\
$^2 p_{3/2} - ^4 s_{1/2}$         &    6.5999         \\
$^4 p_{1/2} - ^4 s_{1/2}$       &      -1.3492        \\
$^4 p_{3/2} - ^4 s_{1/2}$       & 6.2365   \\
$^6 p_{3/2} - ^4 s_{1/2}$       & 7.3533   \\
     \hline
\end{tabular}        
 \caption{ Fine- and hyperfine contributions to the Lamb shift in muonic
   deuterium.   }
  \label{tab:HFS-mu-d}
\end{center}   
\end{table}

Table\,\ref{tab:HFS-mu-He} gives the contributions
to the transition energies due to fine and hyperfine structure in muonic
$^3$He,  relative to the 2s-2p$_{1/2}$ transition energy given in
Table\,\ref{tab:Total-He3}.  
\begin{table}[!h]
\begin{center} 
    \begin{tabular}{|lr|}
     \hline
Transition   &   Energy shift in meV  \\
     \hline
$^1 p_{1/2} - ^3 s_{1/2}$        &   85.439       \\
$^3 p_{1/2} - ^1 s_{1/2}$       &   -139.569        \\
$^3 p_{1/2} - ^3 s_{1/2}$       &    26.806        \\
$^3 p_{3/2} - ^1 s_{1/2}$         &  35.270         \\
$^3 p_{3/2} - ^3 s_{1/2}$         &   201.645         \\ 
$^5 p_{3/2} - ^3 s_{1/2}$        & 177.361   \\
     \hline
\end{tabular}        
 \caption{ Fine- and hyperfine contributions to the Lamb shift in muonic
   $^3$He.   }
  \label{tab:HFS-mu-He}
\end{center}   
\end{table}

\newpage
\subsubsection*{Conclusions }
\vspace{-0.2cm}
Experimental precision has reached the point that some previously
neglected effects should be taken into account.  A few of these have been
discussed in this paper

The consequence of the model dependence of the coefficient of  $r^3$
(in units meV\,fm$^{-3}$) for the determination of the nuclear radius has
been discussed.  The conversion of the coefficient of the third Zemach radius
to $( \langle r^2 \rangle)^{3/2}$ is not unique.  For the case of
hydrogen, the contribution 0.009126\,meV\,fm$^{-3}\, \langle r^3 \rangle_{(2)}$
is most plausibly 0.0365(18)\,meV\,fm$^{-3}( \langle r^2 \rangle)^{3/2}$. 
This will increase somewhat the theoretical error given in \cite{experiment}.  
Assuming the value of the proton radius given in \cite{experiment}, the
contribution to the transition energy would be 0.02179$\pm$0.00107\,meV
(here only the uncertainty in the coefficient was considered).

The radiative correction to the contribution to the energy shift of the
2s-state due to nuclear size has been included and the nuclear size
contribution to the vacuum polarization correction has been calculated
more completely.  Details are given in Appendix~B.

The previously neglected corrections to the muon Lamb shift discussed in
Appendix~C have not been  included in the summaries.  They introduce an
additional theoretical uncertainty, which is of the order of 1-2\,$\mu$eV     
for hydrogen.  They are not negligible for the other cases studied here.  
Although the basic conclusions in the hydrogen experiment are unchanged, 
the theoretical uncertainty is increased, which would slightly increase the
uncertainty in the determination of the proton radius.

The numerical results for muonic helium differ somewhat from those of
Martynenko \cite{martynenko07}.  It is difficult to compare all terms,
but Martynenko's calculation of the Wichmann-Kroll contribution is
incorrect (the sign is wrong, for reasons described in section 9.3.1 of
ref.\,\cite{eides}).  A corrected version \cite{martynenko14} has
appeared very recently.  

The  hyperfine splitting of the 2s-state has been
recalculated, including some previously neglected effects.  The result
for muonic hydrogen is \\
\mbox{(22.9830$\pm0.0027$)\,meV-0.16037\,meV\,fm$^{-1}\langle r \rangle_{(2)}$.}   
\\


For muonic deuterium the total hyperfine splitting of the 2s-state is \\
\mbox{6.17621\,meV-0.045446\,meV\,fm$^{-1}\langle r \rangle_{(2)}$\,=\,6.0584(7)\,meV }
\\

A relativistic nonperturbative calculation of these effects is
desirable, and was  performed in part \cite{carroll}.  Since some
problems with the hyperfine structure of the 2s state have been
mentioned, no further comments will be given here.  The more recent
result of a relativistic nonperturbative calculation of these effects   
in hydrogen \cite{Indelicato} agrees quite well with the results presented 
here. 


\subsubsection*{Acknowledgment }
\vspace{-0.2cm}
The author wishes to thank F. Kottmann   
for extensive email correspondence regarding this work. 

\subsubsection*{Appendix A: Details of the relativistic recoil calculation }
\vspace{-0.2cm}
Although the results of the alternative approach to relativistic recoil
corrections given in ref.\cite{Karshenboim2012} are used in the main
part of this paper, the approach used previously is described here. An
error in that previous work has been corrected.
As mentioned before, the energy levels of muonic atoms are given, to
leading order in $1/m_N$ by
\begin{equation*}
  E~=~E_r - \frac{B_0^2}{2  m_N} + \frac{1}{2 m_N} \langle h(r) +  
  2 B_0 P_1(r) \rangle
\end{equation*}
where $E_r$ is the energy level calculated using  the Dirac equation with
reduced mass and $B_0$ is the unperturbed binding energy.  Also
\begin{equation*}
   h(r)\, = \, - P_1(r)(P_1(r) + \frac{1}{r} Q_2(r))  
            - \frac{1}{3 r} Q_2(r) [P_1(r) + \frac{ Q_4(r)}{r^3}]
\end{equation*}
where  
\begin{align} \label{eq:rmp3}
 P_1(r)&\,=\, 4 \pi \alpha Z \int_r^{\infty} r' \rho(r') dr' 
     & \,=\,&  -V(r)-rV'(r)  \\ \nonumber
Q_2(r)&\,=\,4 \pi \alpha Z \int_0^r r'^2 \rho(r') dr' & \,=\,&  r^2V'(r) \\ \nonumber   
Q_4(r)&\,=\,4 \pi \alpha Z \int_0^r r'^4 \rho(r') dr' & \,=\,& r^4V'(r) - 2 r^3V(r) + 6 \int_0^r r'^2 V(r') dr'     
\end{align}
The calculation of these corrections in first order perturbation theory,
and to first order in the Uehling potential, is described in detail here.  
It was also given in ref.\,\cite{Borie05}.  Some minor mistakes in  the
numerical calculation have been corrected.
In the case of a point Coulomb potential ($V(r)\,=\,-\alpha Z/r $), the
functions $P_1(r)$ and $Q_4(r)$ are identically zero.  The total potential
considered here is  $V(r)\,=\,-\alpha Z/r + V_{Uehl}(r)$.

For the contribution due to the term $B_0^2/2m_N$ the unperturbed
binding energy is taken to be equal to  $B_{0D}+ \langle V_{Uehl}\rangle$, 
where $B_{0D}$ is the point Coulomb Dirac binding energy.  
The contribution to $B_0^2/2m_N$ that is linear in the Uehling potential
is then  $(-B_{0D} \langle V_{Uehl}\rangle)/m_N $. 

Note that 
\begin{equation*}
  P_1(r) + \frac{1}{r} Q_2(r) \, = \, - V(r)
\end{equation*}
Thus it is possible to rearrange $h(r)$ to give 
\begin{equation*}
  h(r)\,=\, P_1(r) V(r) - \frac{1}{3 r} Q_2(r) [P_1(r) + \frac{ Q_4(r)}{r^3}]
\end{equation*}


As noted elsewhere in this paper  (see Eq.(\ref{eq:epsvp})), 
an effective  effective charge density $\rho_{VP}$ for vacuum
polarization  can be derived from the Fourier transform of the Uehling
potential. Recall that (for a point nucleus)
\begin{equation*} \begin{split}
V_{Uehl}(r) & \,=\,-\frac{\alpha Z}{r} \frac{2 \alpha}{3\pi} \cdot 
\chi_1(2 m_e r) \\ & \,=\, -(\alpha Z) \frac{2 \alpha}{3\pi} \cdot 
\int_1^{\infty} dz \frac{(z^2-1)^{1/2}}{z^2} \cdot 
\left(1+\frac{1}{2 z^2}\right) \left( \frac{2}{\pi} 
  \int_0^{\infty} \frac{q^2  \cdot  j_0(qr)}{q^2 + 4 m_e^2 z^2}\,dq\right) 
\end{split} \end{equation*}
\noindent where $\chi_n(x)$ is defined in \cite{RMP}.
In momentum space, the Fourier transform of 
$\nabla^2 V$ is obtained by multiplying the Fourier transform of $V$ by
$-q^2$.  
One then obtains 
\begin{equation} \begin{split}
4 \pi \rho_{VP}(r) & \,=\, \frac{2 \alpha}{3\pi} \cdot 
\int_1^{\infty} dz \frac{(z^2-1)^{1/2}}{z^2} \cdot 
\left(1+\frac{1}{2 z^2}\right) \left(\frac{2}{\pi} 
\cdot \int_0^{\infty} \frac{q^4 \cdot j_0(qr)}{q^2 +4 m_e^2 z^2}\,dq \right) \\
& \,=\, \frac{2}{\pi} \cdot \int_0^{\infty} q^2 U_2(q)  j_0(qr)  \,dq
\label{eq:rhovp}
\end{split} \end{equation}
\noindent  $U_2(q)$ is defined in \cite{RMP} (see also Eq.\,\ref{eq:f-phi}).   

Keeping only the Coulomb and Uehling potentials, one finds 
\begin{align*}
 P_1(r) &\,=\,- \alpha Z \frac{2 \alpha}{3\pi} (2 m_e) \chi_0(2 m_e r) \\
 Q_2(r) & \,=\, \alpha Z \left(1 + \frac{2 \alpha}{3\pi}[\chi_1(2 m_e r)
 +  (2 m_e r) \chi_0(2 m_e r)] \right)  \\
 Q_4(r) &\,=\,  \alpha Z \frac{2 \alpha}{3\pi} 
\int_1^{\infty} dz \frac{(z^2-1)^{1/2}}{z^2}  
\left(1+\frac{1}{2 z^2}\right) \\   & \cdot \left( \frac{2}{\pi} \right) 
 \int_0^{\infty} \frac{1}{q^2 + 4 m_e^2 z^2} \frac{(6qr-(qr)^3)\cos(qr)
   + (3(qr)^2-6)\sin(qr)}{q}  \,dq 
\end{align*}

Since vacuum polarization is assumed to be a relatively small correction
to the Coulomb potential, it will be sufficient to approximate
$Q_2(r)/r$ by $\alpha Z/r $   and  \\
$ P_1(r)(P_1(r) + \dfrac{1}{r} Q_2(r))= P_1(r) V(r)$ by  $-\dfrac{\alpha Z}{r} P_1(r)$. \\

Since the correction is to be calculated to linear order in the vacuum
polarization potential, it will also be sufficient to use point Coulomb wave
functions to calculate the expectation values.  Also, for this case, one
can use Schroedinger wave functions, since an accuracy of 1\% will be
adequate.  
After some algebra, one can reduce the expectation values to single
integrals: 
\begin{equation*} \begin{split}
\langle  P_1(r) \rangle \,=\, & - 2 m_e \alpha Z \frac{2 \alpha}{3 \pi}
 \int_1^{\infty} \frac{(z^2-1)^{1/2}}{z}\cdot \left(1+\frac{1}{2
 z^2}\right) \cdot \\ & ~~~~~~~~~~~~~~~~~\left(\frac{(az)^2-az+1}{(1+az)^5}  
 \delta_{\ell 0} +\frac{1}{(1+az)^5}\delta_{\ell 1} \right)
 \,dz 
\end{split} \end{equation*}

\begin{equation*} \begin{split}
\langle \frac{\alpha Z}{r} P_1(r) \rangle \,=\,
 &   - (\alpha Z)^3  m_r m_e \frac{2 \alpha}{3 \pi}
 \int_1^{\infty} \frac{(z^2-1)^{1/2}}{z}\cdot \left(1+\frac{1}{2
 z^2}\right) \cdot \\ & ~~~~~~~~~~~~~~~~~\left(\frac{2(az)^2+1}{2 (1+az)^4}  
 \delta_{\ell 0} +\frac{1}{2(1+az)^4}\delta_{\ell 1} \right)
 \,dz 
\end{split} \end{equation*}
\noindent  with $a \,=\,2 m_e /(\alpha Z m_r)$.  

Finally,
\begin{equation*} \begin{split}
\langle \frac{\alpha Z}{3 r^4} Q_4(r) \rangle \,=\,
 &  \frac{(\alpha Z)^4  m_r^2}{6} \frac{2 \alpha}{3 \pi}  \int_1^{\infty}
 \frac{(z^2-1)^{1/2}}{z^2}\cdot \left(1+\frac{1}{2 z^2}\right) \cdot \\
 & ~~~~~~~~~~~~~~~\Biggl[ \Bigl[-\frac{6}{az}
 \Bigl(\frac{2 +az}{1+az}-\frac{2}{az} \ln(1+az) \Bigr)  \\
  &~~~~~~~~~~~~~ + \frac{az(8(az)^2+9az+4}{2(1+az)^4} \Bigr]  
 \delta_{\ell 0} - \frac{(az)^2}{2(1+az)^4} \delta_{\ell 1} \Biggr]
 \,dz 
\end{split} \end{equation*}

Combining these expectation values according to equations \ref{eq:rmp1}
and \ref{eq:rmp2}, one finds a contribution to the 2p-2s transition of 
-0.00348\,meV (hydrogen) and -0.00208\,meV (deuterium), as well as
-0.0262\,meV for $^3$He and -0.0203\,meV for $^4$He.  
To obtain the full relativistic and recoil corrections,
one must add the difference between the expectation values of the
Uehling potential calculated with relativistic and nonrelativistic wave
functions.   The  
previous value of the total correction of for muonic hydrogen was 0.0169\,meV,
 in very good agreement with the correction of .0168\,meV calculated by 
Veitia and Pachucki \cite{Pachuki3}.  The treatment presented here has the
advantage of treating the main contribution relativistically and applying
a small correction that can be calculated using first order perturbation
theory. 

A similar relativistic recoil correction  for finite nuclear size should
be included in a relativistic calculation starting from the Dirac
equation (see Ref.\,\cite{friar79}).  However it is automatically taken
into account when the calculation is based on a generalized Breit
equation, such as in Ref.\,\cite{Pachuki3}.

\subsubsection*{ Appendix B:  Higher order contributions to the
  correction for finite nuclear size }
\vspace{-0.2cm}
If the transition energy is written in the form 
\begin{equation*}
\Delta E_{LS}\,=\, A + B \langle r^2 \rangle + C (\langle r^2 \rangle)^{3/2}
\end{equation*}
it is necessary to calculate the quantities $A$, $B$, and $C$.
Suggested values for $C$ have been given in the text, and $A$ can be
determined from the summary tables. 
Here the higher order contributions to $B$ 
mentioned previously are explicitly calculated for all four cases.
There are several contributions and one finds 
\begin{equation*}
B\,=\, b_{a} + b_b + b_c + b_d + b_e 
\end{equation*}

 From Eq.\,\ref{eq:FS-friar}, the main term in the expression for the
energy shift of the 2s-state is 
\begin{equation*}
\Delta E_{ns}\,=\,-\frac{2 \alpha Z}{3} \left(\frac{\alpha Z m_r}{n}\right)^3
  \langle r^2 \rangle \,=\,b_{a} \langle r^2 \rangle
\end{equation*}
This defines $b_{a}$.

Radiative corrections to the main term have been calculated by Eides and
Grotch \cite{eides-grotch}.  They contribute an additional correction
of order $\alpha (\alpha Z)^5$, which amounts to 
\begin{equation*}
-\frac{2 \alpha Z}{3} \left(\frac{\alpha Z m_r}{n}\right)^3
  \alpha^2 Z (23/4 -4 \ln(2) -3/4)  \langle r^2 \rangle\,=\,
 \alpha^2 Z (5 - 4 \ln(2)) \cdot b_{a} \langle r^2 \rangle\,=\,b_{b} \langle r^2 \rangle
\end{equation*}

Next, more details for the evaluation of the order $(\alpha Z)^6$  
contributions to the correction for finite nuclear size to the energy of
the 2s~state are given.   

The terms given by Friar\,\cite{friar79} involving 
$(\alpha Z)^2 (F_{REL}+m^2_r F_{NR})$ 
in equation\,(\ref{eq:FS-friar}) are given for n=2 explicitly by 
\begin{equation*}
 F_{REL} \,=\, -\langle r^2 \rangle [\gamma - \frac{35}{16} + \ln (\alpha Z)
  + \langle \ln(m_r r) \rangle]
  - \frac{1}{3}\langle r^3 \rangle \langle\frac{1}{r} \rangle 
  + I_2^{REL} + I_3^{REL}    
\end{equation*}%
\noindent and 
\begin{equation*}
 F_{NR} \,=\, \frac{2}{3} (\langle r^2 \rangle)^2 [\gamma - \frac{5}{6}
 + \ln (\alpha Z)] + \frac{2}{3}\langle r^2 \rangle \langle r^2 \ln(m_r r) \rangle  
 + \frac{\langle r^4 \rangle}{40} + \langle r^3 \rangle \langle r \rangle
 + \frac{1}{9}\langle r^5 \rangle \langle\frac{1}{r} \rangle
 + I_2^{NR} + I_3^{NR}    
\end{equation*}%
The expectation values for the various moments of the charge
distribution are given in Appendix~E of ref.\,\cite{friar79}.
Approximate values for the integrals 
 $I_2^{REL}$,  $I_3^{REL}$,  $I_2^{NR}$ and $I_3^{NR}$ are given only
for uniform and exponential charge distributions, so these extra terms,
 with the exception of the contribution to $ F_{REL}$ that are
 proportional to $\langle r^2 \rangle$ were calculated for an
 exponential charge distribution, even if this is not completely
 realistic.  The additional contribution proportional to  
$\langle r^2 \rangle$ in  the finite size correction to the 2s level in
 equation\,(\ref{eq:FS-friar}) is  
\begin{equation*}
 \frac{2 (\alpha Z)^3}{3} \left(\frac{\alpha Z m_r}{2}\right)^3 
 \langle r^2 \rangle [\gamma - \frac{35}{16} + \ln (\alpha Z)]\,=\,b_{c} \langle r^2 \rangle
\end{equation*}%
The remaining terms are small, but a model independent evaluation is
prohibitively difficult.  Numerical values are given in Tables (3,4,5,6)
with the heading "remaining order  $(\alpha Z)^6$".

There are two corrections due to finite nuclear size to the vacuum
polarization contribution.  One of these was obtained as a result of
numerical integration of the expectation value of the Uehling potential,
as discussed previously.  It will be denoted by  $b_{d} \langle r^2 \rangle$
and numerical results are listed in Table\,\ref{tab:coeff}.  The (small)
correction arising from the K\"allen-Sabry potential was also included.
The other correction, 
due to iterations, has been calculated for hydrogen by Eides et
al.\,\cite{eides} and by Pachucki\,\cite{Pachuki1}.  It is given by  
\begin{equation*}
\Delta E_{2s}\,=\,-\frac{4 \pi \alpha Z}{3} \langle r^2 \rangle
 \int V_{Uehl}(r) \psi_{2s}(r) G_{2s}'(r,0) \psi_{2s}(0) d^3r
\end{equation*}
Ivanov and Karshenboim \cite{karshenboim3}  have given an expression for 
$ G_{2s}'(r,0)$.  With \mbox{$x=\alpha Z m_r r = b r $,} it is  
\begin{equation*}
    G_{2s}'(r,0) \,=\, \frac{\alpha Z m^2_r}{4 \pi} \frac{ e^{-x/2}}{2x} 
\left[4x(2-x)(\ln(x)+\gamma)+ 13x^2-6x-x^3 -4 \right]
\end{equation*}
\noindent ($\gamma$ is Euler's constant). Recall that 
\begin{equation*}
  \psi_{2s}(r)\, =\, \psi_{2s}(0) (1-x/2) \exp(-x/2)
\end{equation*}
Then 
\begin{equation*} 
 \psi_{2s}(r) G_{2s}'(r,0) \psi_{2s}(0) \,=\,\alpha Z m^2_r \frac{b^3}{8 \pi^2} 
(1-x/2) e^{-x}\left[(1-x/2)(\ln(x)+\gamma)+\frac{13x-6-x^2}{8}-\frac{1}{2x}\right] 
\end{equation*}
It is possible to show that the integral is proportional to the
correction $\varepsilon_{VP2}$ to the hyperfine splitting of the 2s-state.
If one rewrites  Eq.(\ref{eq:delta-psi}) in terms of $x=2y$, one finds 
\begin{equation*} \begin{split}
 \psi_{2s}(r) \delta_M \psi_{2s}(r) \,=\, & \frac{b^3}{8 \pi}
  \left(\frac{2}{b}\right)^3 (\alpha Z m^2 \Delta \nu_F)(1-x/2) e^{-x} \\
& ~ \left[(1-x/2)(\ln(x)+\gamma)+\frac{13x-6-x^2}{8}-\frac{1}{2x}\right]
\end{split}
\end{equation*}
\noindent Thus
\begin{equation*}
  \psi_{2s}(r) G_{2s}'(r,0) \psi_{2s}(0) \,=\, \frac{b^3}{8\pi} 
  \frac{1}{\Delta \nu_F }\psi_{2s}(r) \delta_M \psi_{2s}(r),
\end{equation*}
and hence
\begin{equation*}
\Delta E_{2s}\,=\,-\frac{4 \pi \alpha Z}{3}\langle r^2 \rangle
\frac{b^3}{8\pi} \frac{1}{\Delta \nu_F }
\int V_{Uehl}(r) \psi_{2s}(r) \delta_M \psi_{2s}(r) d^3r.  
\end{equation*}
By definition, the quantity $\varepsilon_{VP2} $ is given by 
\begin{equation*}
\varepsilon_{VP2} \,=\,\frac{2}{\Delta \nu_F } \int V_{Uehl}(r)
\psi_{2s}(r) \delta_M \psi_{2s}(r) d^3r. 
\end{equation*}
Hence 
\begin{equation*}
\Delta E_{2s}\,=\,-\frac{ \alpha Z}{12} b^3 \langle r^2 \rangle
 \varepsilon_{VP2} \,=\, b_{e} \langle r^2 \rangle 
\end{equation*}
Since the two-loop VP corrections to  $\varepsilon_{VP2} $ have been
calculated, they are included in $b_e$.  Effectively, the one-loop
correction is multiplied by approximately 1.0077.   
Table\,\ref{tab:coeff} gives numerical values of these coefficients (in
meV\,fm$^{-2}$).   The total for hydrogen differs slightly from that given 
in ref.\,\cite{experiment} (-5.2262meV\,fm$^{-2}$).  The difference is due
partly to a more precise determination of the coefficient  $b_{c}$  and
partly to the inclusion of the radiative correction ($b_{b}$).  
\begin{table}[!h]
\begin{center} 
    \begin{tabular}{|lrrrr|}
      \hline
          &  Hydrogen  &   deuterium  &  $^3$He&  $^4$He \\
 $b_{a}$     &   -5.1973~   & -6.0730~   & -102.520~   &  -105.319~    \\
 $b_{b}$     &    - 0.00062    & -0.00072 & -0.0243  & -0.0250 \\
 $b_{c}$     &    - 0.00181    & -0.00212  & -0.1275  & -0.1310 \\
 $b_{d}$     &    - 0.0110~    & -0.0130~   & -0.3176   & -0.3297 \\
 $b_{e}$     &    - 0.0165~    & -0.02062  & -0.5217  & -0.5392 \\
\hline
$B_{2s}$=total (2s) & -5.22723    & -6.10946 & -103.511~  &  -106.344~ \\

  $b(2p_{1/2}) $    &  -0.0000519   & -0.0000606   &-0.00409  & -0.004206   \\
\hline
\end{tabular}        
 \caption{ Contributions to the coefficient of  $\langle r^2 \rangle$
  for the energy shift  of the 2s and 2p-states in muonic
   hydrogen, deuterium, $^3$He and $^4$He, in  meV\,fm$^{-2}$.  }
  \label{tab:coeff}
\end{center}   
\end{table} 





It is interesting to note that the total coefficient of 
$\langle r^2\rangle$ calculated here for  $^4$He is in excellent
agreement with the calculated value given in ref.\,\cite{Borie78}, where    
the finite size correction for  $^4$He was estimated to  
be -106.2\,$\langle r^2 \rangle$\,+\,1.4\,$\langle r^3 \rangle$ (in meV).
 An additional term proportional to the Zemach radius is discussed in the
section on relativistic recoil.  For $^4$He it was parametrized as   
0.4\,$\langle r \rangle$.  It is actually proportional to 
$\langle r \rangle_{(2)}$, and was evaluated in the section on
relativistic recoil. 

For $^4$He the
binding energy of the  $2p_{1/2}$-state is decreased by 0.0148\,meV (for
a radius of 1.676\,fm and a Gaussian charge distribution, which is a
fairly good approximation for helium, according to Friar \cite{friar79}).  
The contribution to the fine structure is 0.0118\,meV.  The difference
is due to the term proportional to $\langle r^4 \rangle$, which is the
same  for both 2p-levels.  If the radius is 1.681\,fm both of these
numbers are increased by 0.0001\,meV.
For $^3$He the binding energy of the  $2p_{1/2}$-level is decreased by
0.0213\,meV (for a radius of 1.966\,fm).  The contribution to the fine
structure is 0.0158\,meV.  

For muonic deuterium, the first term of
equation\,(\ref{eq:FS-friar})would  contribute   \\
-6.0730\,$\langle r^2 \rangle$\,=\,-(27.817$\pm$0.078)\,meV (using the
value of the radius from the newest CODATA compilation \cite{codat06}).  
If the radius of 2.130(3) from ref.\,\cite{sick98} is used, this 
contribution is -26.8585$\pm$.076meV.  Using the total coefficient from 
 Table\,\ref{tab:coeff} gives  -27.718$\pm$.076meV.

 The term proportional to $\langle r^3 \rangle_{(2)}$
(the coefficient is 0.0112\,meV\,fm$^{-3}$) gives a contribution of
0.382\,meV or 0.417\,meV, depending on the model for the charge density,
and a radius of 2.14\,fm.  In ref.\cite{codat06} it is suggested to
calculate this term according to the prescription 
\begin{equation*}
 \langle r^3 \rangle_{(2)} \,\approx\,4.0(0.2) (\langle r^2 \rangle)^{3/2}
\end{equation*}
Using this value, one obtains 0.439(22)\,meV with a radius of 2.14\,fm
(0.433(22)\,meV with a radius of 2.13\,fm).  
The last terms in Eq.(\ref{eq:FS-friar}) contribute a term proportional
to $\langle r^2 \rangle$ and remaining terms.  The first is 
\mbox{$b_{c} \langle r^2 \rangle$\,=\,-0.00966\,meV;} and the remaining 
terms (of order $(\alpha Z)^6$) given in \cite{friar79}  contribute 
0.00337\,meV. 
The complete contributions are given in Table\,\ref{tab:Total-D}.

\subsubsection*{ Appendix C:  Further corrections }
\vspace{-0.2cm}

A number of the corrections described here, in particular the "muon Lamb
shift" and the hyperfine structure of s-states, involve the expectation
value of  $\nabla^2 V$.  Note that using the normalizations of
\cite{RMP,hyperfine}, one has $\nabla^2 V = - 4 \pi \alpha Z \rho$ where
$\rho$ is the nuclear charge density.  

Usually $\nabla^2 V\,=\,-4\pi \alpha Z \rho(r)$ is approximated by a
delta function, giving $V\,=\,- \alpha Z/r$ and 
\begin{equation*}
\langle\nabla^2 V\rangle\,=\,-4\pi \alpha Z |\psi_{ns}(0)|^2 \delta_{\ell 0}
\,=\,-4 Z \alpha \big(\frac{Z \alpha m_r}{n}\big)^3 \delta_{\ell 0}
\end{equation*}
 However, the potential should be corrected for vacuum polarization due
 to electron loops, and, at least for s-states, for the effect of finite
 nuclear size.  This has been done (a long time ago)
 \cite{hyperfine,Brodsky,sternheim,zemach} for the hyperfine
 structure,  but up to now not for the  "muon Lamb shift". 

Calculations of the correction due to a vacuum polarization insert in
the external photon line for hydrogen by Pachucki \cite{Pachuki1} and 
 by Jentschura \cite{jentschura} do not agree.  The contribution is
estimated to be at least comparable to the experimental uncertainty, so
this requires further investigation.  
The energy shifts for at least some of the terms in this contribution
as calculated by  Jentschura can be compared directly. 
Here  vacuum polarization corrections to the Bethe logarithms will not
be calculated, but  
vacuum polarization corrections to the expectation values of  
$\nabla^2 V$ and of $\dfrac{1}{r} \dfrac{d V}{dr}$ can be calculated
quite easily, and a detailed numerical comparison for these parts is possible.  

The largest correction to the muon self-energy graphs involves the 
expectation value of  $\nabla^2 V$.  
An  effective charge density $\rho_{VP}$ for vacuum
polarization  can be derived from the Fourier transform of the Uehling
potential, and was given by Eq.\,(\ref{eq:rhovp}). 
It is straightforward to calculate the expectation value of 
$4 \pi \rho_{VP}(r)$.  The result for hydrogen is \\
$\langle \nabla^2 V_{VP} \rangle_{2s}\,=\,4.4617\times 10^{-11}\,fm^{-3}$ 
and $\langle \nabla^2 V_{VP} \rangle_{2p}\,=\,-8.7788\times 10^{-13}\,fm^{-3}$. 
As a check on this calculation, the correction to the
contribution of a muon loop in the external photon line to the "muon
Lamb shift" was calculated directly from the expectation value of 
$4 \pi \rho_{VP}(r)$.  It gave precisely one half of the
contribution normally labeled "mixed vacuum polarization", which would
be expected.  This number has been checked independently by other
authors (\cite{eides}).  The factor 2 comes from the fact that the two
loops can appear in either order.  The value of 
$\langle \nabla^2 V_{VP}\rangle_{2s}$  turns out to be equal to  
$\varepsilon_{VP1}$ (see Eq.\,\ref{eq:epsvp}) times the point Coulomb value. 

An additional correction due to distortion of the wave function of the
2s-state,  was given  by Ivanov et al. \cite{ivanov} (see also
\cite{karshenboim-hfs}), and corresponds to 
multiplying the point Coulomb value by $\varepsilon_{VP2}$, which, as
mentioned before, has the same numerical value.  Strictly speaking, this
correction does not correspond to a  vacuum polarization insert in
the external photon line, since it involves a second external photon,
but it is numerically just as important as $\varepsilon_{VP1}$.  

The main contribution to the muon Lamb shift has been given by 
(\cite{eides,Pachuki1})
\begin{equation*}
 \Delta E_{LS} \,=\, \frac{4 Z \alpha^2}{3 \pi m^2} \big(\frac{Z \alpha
   m_r}{n}\big)^3 \Big[ \big[\ln\big(\frac{m}{(Z \alpha)^2 m_r}\big) +
 \frac{11}{24} + \frac{3}{8}\big] \delta_{\ell 0} - \ln k_0(n,l) \Big]
\end{equation*}
Rewriting this in terms of the expectation value of $\nabla^2 V$, and
including the contribution of the lowest order anomalous magnetic moment
that is proportional to $\nabla^2 V$ results in 
\begin{equation*}
 \Delta E_{LS} \,=\, \frac{\alpha}{3 \pi m^2} \langle\nabla^2 V\rangle
 \Big[ \big[\ln\big(\frac{m}{(Z \alpha)^2 m_r}\big) + 
 \frac{5}{6} \big]  - \ln k_0(n,l) \Big]
\end{equation*}
In this work, corrections to the Bethe logarithm will not be  calculated. 
The contributions that can  definitely be compared with
ref.\,\cite{jentschura} are then
\begin{equation}
 \Delta E_{LS,VP} \,=\, \frac{\alpha}{3 \pi m^2} \langle\nabla^2 V_{VP}\rangle
 \Big[\ln\big(\frac{m}{(Z \alpha)^2 m_r}\big) + 
 \frac{5}{6}  \Big]
\label{eq:UJ-ls}
\end{equation}
Calculating the energy shifts with the values calculated from
Eq.\,\ref{eq:UJ-ls} for  
$\langle \nabla^2 V_{VP} \rangle$ gave energy shifts of 0.00191\,meV for 
the 2s~state and -0.00004\,meV for the $2p_{1/2}$~state.  Including the
contribution from $\varepsilon_{VP2}$ increases the shift for
the 2s-state to 0.00486\,meV  for  a total
contribution of -0.00490\,meV.  Including the effect of finite size on
the Uehling potential reduced the number for the 2s~state by 0.00006\,meV.

In Eqs.\,(3.6) and (3.14) of Ref.\,\cite{jentschura} the 10/9 should be
replaced by 5/6 in order to ensure consistency with Eq.\,(2.28).  Doing
that, and calculating the expression for the same pieces of Eq.\,(3.14)
as in Eq.\,\ref{eq:UJ-ls},  which is 
\begin{equation*}
 \Delta E_{LS,VP} \,=\, \frac{Z \alpha^3}{\pi^2 m^2}  
\big(\frac{Z \alpha  m_r}{n}\big)^3  V_{61} 
 \Big[ \ln\big(\frac{m}{(Z \alpha)^2 m_r}\big) + 
 \frac{5}{6}  \Big]
\end{equation*}
with $ V_{61} $\,=\,3.09 for the 2s~state and $ V_{61} $\,=\,-0.023 for
the 2p$_{1/2}$~state gives energy shifts of 0.004885\,meV  for the
2s~state and -0.000036\,meV for the $2p_{1/2}$~state.  These numbers
agree fairly well with those calculated here.

The contribution to the spin-orbit shift due to the muon's anomalous
magnetic moment, with $a_{\mu}=\alpha/2\pi$ can also be checked.  
The expression for  $\dfrac{1}{r} \dfrac{d V}{dr}$ including the Uehling
potential is given  by Eq.\,(\ref{eq:vp2p}).  
 $\varepsilon_{2p}$ was defined as the ratio of
 the value of the expectation value for the Uehling potential to that
for the point Coulomb potential ($\alpha Z/r^3$), and  an expression is
given in  Eq.\,(\ref{eq:eps2p}).  The numerical integration is
straightforward and has been checked for accuracy.  The result for
hydrogen is  $\varepsilon_{2p}$\,=\,0.000365.  Thus, to obtain the
energy shift of the $2p_{1/2}$ state, one simply has to multiply the
standard value \mbox{(-0.011713\,meV)} by  $\varepsilon_{2p}$, giving  
$-4.275\times 10^{-6}$\,meV.  Evaluating Jentschura's Eq.(3.10), which
neglected a factor  $m_{\mu}/m_r$,  gave $-3.2265\times 10^{-6}$\,meV.  
Correcting for the missing mass factor gave $-3.590\times 10^{-6}$\,meV.  

For an estimate of the effect of nuclear size, an
exponential charge density is used, mainly because all integrals can
then be calculated analytically.  The charge density is then
\begin{equation*}
 \rho (r) \,=\, - \frac{1}{8 \pi r_0^3 } e^{-r/r_0}   
\end{equation*}%
Evaluation of the expectation value gives (expanded to lowest order in
$(\alpha Z m_r r_0)$) 
\begin{equation*}
 \langle \nabla^2 V_C \rangle_{2s} \,=\,
 - \frac{(\alpha Z)^4 m^3_r}{2} [1-6(\alpha Z m_r r_0)+21(\alpha Z m_r r_0)^2] 
\end{equation*}%
and 
\begin{equation*}
 \langle \nabla^2 V_C \rangle_{2p} \,=\,
 - \frac{(\alpha Z)^4 m^3_r}{2} (\alpha Z m_r r_0)^2 
\end{equation*}%
Note that $\alpha Z m_r r_0 \approx 0.00087$  for muonic hydrogen,
giving a value for $(6 \alpha Z m_r r_0)$ of 0.0052.  As was
demonstrated a long time ago \cite{zemach}, the correction due to finite
nuclear size is probably given correctly by 
$2 \alpha Z m_r \langle r\rangle_{2}$ (here with charge density rather
than with magnetization density).   The expectation
value of  $\nabla^2 V$ for the 2p-state is proportional to  
$(\alpha Z m_r r_0)^2\approx 7.7 \cdot 10^{-7}$, and is thus much smaller.
However, one finds (to linear order in  $\alpha Z m_r r_0$)
\begin{equation*}
 \Big\langle \frac{1}{r}\,\frac{dV}{dr} \Big\rangle \,=\,
  \frac{\alpha Z (\alpha Z m_r)^3}{24} (1 - \frac{9 \alpha Z m_r r_0}{2}) 
\end{equation*}%
The finite size correction to the muon self-energy of the 2s-state is then
approximately 0.00699\,meV for a Zemach radius of 1.086\,fm for hydrogen. 
The total correction to the muon self energy contribution (which is
negative) is then 0.00207\,meV,  which is too small to explain the
discrepancy in the measurement of the proton radius.


 
Since all of these corrections have also been calculated for deuterium
and the helium isotopes, one can give a set of partial corrections also
for these cases.  Numerical values for the corrections  
\mbox{$\varepsilon_{VP1} + \varepsilon_{VP2}$} and $\varepsilon_{Zem}$ for the
four cases discussed here are given in Table \ref{tab:correction}. 
\begin{table}[!h] 
\begin{center} 
    \begin{tabular}{|lrr|}
   \hline
Hydrogen           & 0.00537  & -0.00762      \\
Deuterium          & 0.00555   &  -0.01918       \\
$ ^3$He            &  0.00821  & -0.03619     \\
$ ^4$He            &  0.00824   & -0.03315   \\
   \hline
\end{tabular}        
 \caption{$(\varepsilon_{VP1} + \varepsilon_{VP2})$ and $ \varepsilon_{Zem}$}
  \label{tab:correction}
\end{center}   
\end{table}    
 

Corrections to the energy shift of the 2s-state due to these effects are
summarized in Table\,\ref{tab:totcorr}. All energies are given in meV.  
The standard value is calculated with Eq.\,(\ref{eq:UJ-ls}), and does not
include all contributions, since 
 corrections to the term involving the Bethe logarithm are not included.
The effect is significant for the helium isotopes.  These corrections were not
included in Table\,\ref{tab:muonLS}. 
\begin{table}[!h] 
\begin{center} 
    \begin{tabular}{|lrrrr|}
   \hline
  &  standard value  & correction (VP) & correction (size) & corrected value  \\
   \hline
Hydrogen        & -0.91689  & -0.004924  & 0.00699  &  -0.91480 \\
Deuterium       & -1.06116  & -0.00589~ &  0.02035  &  -1.04670  \\
$ ^3$He         & -15.5515~  &-0.1277~~ &  0.5630~  &  -15.1162~   \\
$ ^4$He         & -15.9455~ &  -0.1314~~ &  0.5294~  &  -15.5475~   \\
   \hline
\end{tabular}         
\caption{total corrections to the muon self energy energy shift of the 2s-state}
 \label{tab:totcorr}
\end{center}   
\end{table}    

As mentioned before, the spin-orbit terms in the "muon Lamb shift"
should also be corrected for  vacuum polarization using the correction
calculated in Eq.\,(\ref{eq:eps2p}) for the  2p-states.
For hydrogen,  $\varepsilon_{2p}$\,=\,0.000365,  for deuterium, 
$\varepsilon_{2p}$\,=\,0.000391, for $^3$He, 
$\varepsilon_{2p}$\,=\,0.000894,  and for    $^4$He, 
$\varepsilon_{2p}$\,=\,0.000902.   
The fine structure splitting would be increased by 0.00001\,meV in
hydrogen and by 0.0003\,meV in $^4$He.  

\subsubsection*{ Appendix D: Relativistic calculation of the hyperfine structure of the 2s state }
\vspace{-0.2cm}
A recent paper by Carroll et al. \cite{carroll} calculates the Lamb
shift in muonic hydrogen relativistically and nonperturbatively. 
Their calculation of the magnetic hyperfine structure of the 
 2s-state  approximates the magnetization density as a delta function. 
 However the magnetization
density has a finite extent, and  vacuum polarization has an effect 
on the magnetic vector potential.  To estimate these effects
relativistically, they will be calculated here with point Dirac wave functions.
Carroll et al. do take into account perturbation  of the wave function,
which is also important.  A completely correct calculation takes both
contributions into account.  

According to Eq.\,(40) of ref.\,\cite{RMP}, the hyperfine structure of an
ns-state  is given by 
\begin{equation}
\Delta E_{n \kappa}~=~\frac{4 \pi \kappa g}{\kappa^2-1/4} \frac{\alpha}{2  m_N}\,
  [F(F+1)-s_2(s_2+1)-3/4]
\cdot \int_0^{\infty} \frac{dr}{r^2} F_{n \kappa}(r)G_{n \kappa}(r)
  \cdot \int_0^{r} dr_N r^2_N \mu(r_N) 
\label{eq:HFS-pert1}
\end{equation}
where $F_{n \kappa}$ and $G_{n \kappa}$ are the small and large components
of the wave function.  $n$ is the principle quantum number and 
 $\kappa\,=\,\ell(\ell+1)-j(j+1)-1/4$.
Nuclear magnetic moments are given in units of nuclear magnetons, so
that the proton has $g=2(1+\kappa_p)$.    
Here the magnetization density is normalized to \\
\qquad  \qquad $ 4\pi \int_0^{\infty} dr_N r^2_N \mu(r_N)\,=\,1 $.  

The  paper by Carroll et al. \cite{carroll} uses the approximation 
$ 4\pi \int_0^{r} dr_N r^2_N \mu(r_N)\,=\,1 $.  However the
magnetization density has contributions from finite extent and from
vacuum polarization that must be taken into account.  Here this will be
cone perturbatively, using   point Dirac wave functions.
The wave functions are given in the book of 
Akhiezer and Berestetskii \cite{akhiezer}. 

For an estimate of the effect of a finite extent on the  magnetization
density an exponential magnetization charge density is used, mainly
because all integrals can then be calculated analytically.  The density is then
\begin{equation*}
 \mu (r) \,=\, - \frac{1}{8 \pi r_0^3 } e^{-r/r_0}   
\end{equation*}%
Here $r_0$ corresponds to the magnetic rms radius of the nucleus, and is
not necessarily the same as that corresponding to the charge radius.

Vacuum polarization affects the magnetic vector potential in the same
manner as it does the electrostatic potential, and the contribution to
the magnetization density is thus essentially
the same as Eq.\,\ref{eq:rhovp}.  
\begin{equation*} \begin{split}
4 \pi \rho_{VP}(r) & \,=\, \frac{2 \alpha}{3\pi} \cdot 
\int_1^{\infty} dz \frac{(z^2-1)^{1/2}}{z^2} \cdot 
\left(1+\frac{1}{2 z^2}\right) \left(\frac{2}{\pi} 
\cdot \int_0^{\infty} \frac{q^4 \cdot j_0(qr)}{q^2 +4 m_e^2 z^2}\,dq \right) \\
& \,=\, \frac{2}{\pi} \cdot \int_0^{\infty} q^2 U_2(q) G_M(q) j_0(qr)  \,dq
\end{split} \end{equation*}
\noindent  where $U_2(q)$ is defined in \cite{RMP}.  See also
Eq.\,(\ref{eq:f-phi}).  The finite extent
of the magnetization  density can be taken into account when the
momentum space representation is used.  As usual, $G_M$ is the magnetic
form factor.  
One then finds
\begin{equation}  \begin{split}
 4\pi \int_0^{r} dr_N r^2_N \mu(r_N)\,=\, & 1 - e^{-r/r_0} \Bigl(1 - \frac{r}{r_0}   
 - \frac{1}{2} (\frac{r}{r_0})^2 \Bigr) \\  
 &  +~ \frac{2 \alpha}{3 \pi} \int_1^{\infty} 
  \frac{(z^2-1)^{1/2}}{z^2}\cdot \left(1+\frac{1}{2 z^2}\right) \cdot
   (1 + 2 m_e r z) \cdot e^{-2 m_e r z} \,dz   
\end{split}
\label{eq:HFS-pert2}
\end{equation}%

If one uses the momentum space representation, one finds  for the vacuum
polarization contribution  
\begin{equation*}
 4\pi \int_0^{r} dr_N r^2_N \rho_{VP}(r_N)\,=\, \frac{2}{\pi} \int_0^{\infty} 
   \left( \frac{\sin(qr)}{q} - r \cos(qr) \right) U_2(q) G_M(q) \,dq
\end{equation*}%
The expression given in  Eq.\,\ref{eq:HFS-pert2} can be derived directly
from this form if $G_M=1$.

It is interesting to notice that if one were to define a "magnetic" potential
 according to   
$\nabla^2 V_M\,=\,4\pi \mu(r)$, then  
\begin{equation*}
 4\pi \int_0^{r} dr_N r^2_N \mu(r_N)\,=\, r^2 \dfrac{d V_M}{dr}
\end{equation*}%
This permits a more transparent comparison with results obtained by
other methods.  It is easy to see that the vacuum polarization
contribution to  Eq.\,\ref{eq:HFS-pert2} is identical to the vacuum
polarization contribution to $Q_2(r)$ given in Appendix~A. 

If the  HFS given by Eq.\,\ref{eq:HFS-pert1} is calculated using 
$ 4\pi \int_0^{r} dr_N r^2_N \mu(r_N)\,=\,1$, corresponding to the first 
term in the expression for a finite magnetization density 
(Eq.\,\ref{eq:HFS-pert2}), for the 2s-state with j=1/2 and $\kappa\,=\,-1$, 
the result is      
\begin{equation}
\Delta E_{2s}\,=\, C_{2s}  \Bigl[-\frac{N_1(N_1+2)}{2\gamma-1} + 
 \frac{N_1^2(N_1+1)^2}{(2\gamma+1)^2} \Bigr] 
\label{eq:HFS-pert3}
\end{equation}
The factor $C_{2s}$ is defined as 
\begin{equation*}
  C_{2s}\,=\,- \frac{\alpha g}{6 m_p} \frac{m_r}{m_{\mu}} \frac{b^2}{4N_1(N_1+1)}
  \frac{2\gamma+1}{\Gamma(2\gamma+1)} \bigl(\frac{(1-\gamma)}{2}\bigr)^{1/2}
 \, [F(F+1)-3/2]
\end{equation*}
with  $\gamma\,=\,\sqrt{1-(\alpha Z)^2}$,  
$N_1\,=\,\sqrt{2(1+\gamma)}$, and  $b\,=\,2\alpha Z m_r/N_1$. 
A factor $m_r/m_{\mu}$ has been included to account for the
fact that the magnetic moment of the muon is defined in terms of the
free space mass, and not the reduced mass.    It was pointed out in   
ref.\,\cite{RMP} that the analysis given there does not take mass
corrections to the hyperfine structure into account.  Similar mass corrections
were made by Carroll et al.  \cite{carroll}. 

If this is expanded in powers of $\alpha Z$, and only the leading
nonvanishing terms are retained, the result is 
\begin{equation*}
\Delta E_{2s}\,\approx\,- \frac{\alpha^2 Z g}{12 m_{\mu} m_p} (2\alpha Z m_r/n)^3
 \, [F(F+1)-3/2]
\end{equation*}
Comparing this with Eq.\,\ref{eq:beta}, it is easy to see that the factor
 multiplying $ [F(F+1)-3/2]$ is $-\beta/2$  if Z=1 and  $g=2(1+\kappa_p)$. 
Thus, except for the missing factor $1+a_{\mu} $, this is nearly the same as
the standard perturbative result.  Numerically, the value for the total
hyperfine splitting calculated here is 22.8079\,meV;  this is equal to
 $(1+ \varepsilon_{Breit}) \beta$ where  $\beta$  is given by
Eq.\,\ref{eq:beta}, as expected, since the Breit correction was defined
in terms of an expansion of the expression for $\Delta E_{2s}$ given in
Eq.\,\ref{eq:HFS-pert3} in powers of $(\alpha Z)^2$, as pointed out in 
ref.\,\cite{eides}.   Carroll et al. obtain 22.8229\,meV for
this contribution, which is not corrected for distortion of the wave
functions due to vacuum polarization or finite nuclear size. 
The additional contribution due to distortion of the wave function due
to vacuum polarization calculated here (0.0744\,meV) is in
fair agreement with that calculated by Carroll et al. (0.0747\,meV). 


The magnetization density also has to be corrected for vacuum polarization. 
A calculation of Eq.\,\ref{eq:HFS-pert1} for the vacuum polarization 
correction to equation\,\ref{eq:HFS-pert2} gives for the energy shift  
\begin{equation*}  \begin{split}
\Delta E_{2s} \,=\, & C_{2s} \frac{2 \alpha}{3 \pi}
\int_1^{\infty}  \frac{(z^2-1)^{1/2}}{z^2}\cdot \Bigl(1+\frac{1}{2 z^2}\Bigr) 
\cdot\frac{1}{(1+az)^{2\gamma-1}}    \\ 
 & ~~~~~~~~ \cdot \Bigl[-N_1(N_1+2)\bigl(\frac{1}{2\gamma-1}+
  \frac{az}{1+az}\bigr)  \\ 
 &~~~~~~~~ +\frac{2(N_1+1)^2}{(2\gamma+1)(1+az)} \bigl[1 +  \frac{\gamma}{1+az}
 \bigl(2 az -\frac{1}{2\gamma+1}  - \frac{az}{1+az} \bigr) \bigr] \Bigr] \,dz   
\end{split}
\end{equation*}
\noindent  where $a=2m_e/b$.  
The resulting contribution to the total hyperfine splitting of 0.0481\,meV 
agrees very well with the perturbative value of 0.0482\,meV calculated 
nonrelativistically.   Including the effect of finite size on the vacuum 
polarization contribution to the magnetization density reduced each of the
nonrelativistic perturbative contributions (modification of
magnetization density and perturbation  of the wave function) by
$0.00114$\,meV.  

To estimate the effect of finite extension of the magnetization density,
the integral for the $r_0$-dependent terms in Eq.\,\ref{eq:HFS-pert2} is 
evaluated.   The resulting correction to the 
energy shift can be expanded in powers of $br_0$.  If only the  leading
term is retained, the result is  
\begin{equation*} 
\Delta E_{2s}\,=\, C_{2s} (br_0)^{2\gamma -1} 
 \Bigl[-N_1(N_1+2)\bigl(\frac{1}{2\gamma-1} + 1 +\gamma\bigr)\Bigr] 
\end{equation*}
The correction to the hyperfine structure of the 2s-state  due to finite
extent of the magnetization density is  proportional to 
$ (br_0)^{2\gamma -1}$;  it has  thus been shown to be predominantly 
linear in the radius, with smaller corrections resulting from higher powers 
of $br_0$.  Corrections due to distortion of the wave functions  due to 
the Coulomb potential in a nonrelativistic perturbative calculation were
historically approximated by using the Zemach radius (see ref.\,\cite{eides}).  
In a nonperturbative calculation, a fit to a formula that is linear plus
quadratic in the radius would be a useful approach.  In
any case, the finite extent of the magnetization density has a 
dominant contribution that is linear in the radius parameter.

The value of  $r_0$ is not certain.  One possibility is to take
$r_0=r_{mag}/\sqrt{12}$;  another is to use  \mbox{$r_0 = 8 r_{Zem}/35$}. 
The radius of the 
magnetization density is not necessarily the same as the charge radius.  
The magnetic radius from the Mainz experiment \cite{mainz2010} was
given as  0.777(17)\,fm, which has recently been modified to  0.803(17)\,fm.   
A recent determination by the group at Jefferson Laboratory \cite{jlab}
gives  0.867(20)\,fm.   Numerically, one finds  a
reduction of the total hyperfine splitting of $-0.111$\,meV if the
(revised) magnetic radius  from the Mainz experiment
\cite{mainz2010} is used, and $-0.120$\,meV if the result from Jefferson
Laboratory \cite{jlab} is used.  Both differ from the usual perturbative 
result (approximately -0.17\,meV), for reasons that are unclear.  

In any case, the correction estimated here 
differs significantly from the range of standard values obtained with
a nonrelativistic, perturbative calculation, so that one can only say
with confidence that the correction is dominated by a term linear in the
radius parameter.  

\small
\def\refname{{\normalsize References}}

\end{document}